\documentclass[aps,prb,twocolumn]{revtex4}
\usepackage{graphicx,amsmath,amsfonts,amssymb,esint}

\DeclareMathOperator{\sech}{sech}

\begin{document}
\title{Propagating spin-wave normal modes: A dynamic matrix approach using plane-wave demagnetizating tensors}
\author{Y. Henry}
\email{yves.henry@ipcms.unistra.fr}\affiliation{Institut de Physique et Chimie des Mat\'eriaux de Strasbourg, UMR 7504, CNRS and Universit\'e de Strasbourg, B.P. 43, F-67037 Strasbourg Cedex 2, France}
\author{O. Gladii}
\affiliation{Institut de Physique et Chimie des Mat\'eriaux de Strasbourg, UMR 7504, CNRS and Universit\'e de Strasbourg, B.P. 43, F-67037 Strasbourg Cedex 2, France}
\author{M. Bailleul}
\affiliation{Institut de Physique et Chimie des Mat\'eriaux de Strasbourg, UMR 7504, CNRS and Universit\'e de Strasbourg, B.P. 43, F-67037 Strasbourg Cedex 2, France}
\date{\today}

\begin{abstract}
We present a finite-difference micromagnetic approach for determining the normal modes of spin-waves \emph{propagating} in extended magnetic films and strips, which is based on the linearized Landau-Lifshitz equation and uses the dynamic matrix method. The model takes into account both short range exchange interactions and long range dipole-dipole interactions. The latter are accounted for through plane-wave dynamic demagnetization factors, which depend not only on the geometry and relative positions of the magnetic cells, as usual demagnetization factors do, but also on the wave vector of the propagating waves. Such a numerical model is most relevant when the spin-wave medium is spatially inhomogeneous perpendicular to the direction of propagation, either in its magnetic properties or in its equilibrium magnetic configuration. We illustrate this point by studying surface spin-waves in magnetic bilayer films and spin-waves channelized along magnetic domain walls in perpendicularly magnetized strips. In both cases, dynamic dipolar interactions produce non-reciprocity effects, where counter-propagative spin-waves have different frequencies.
\end{abstract}

\maketitle

\section{Introduction\label{Sec_Introduction}}
With the increasingly fast development of magnonics, which is the research field devoted to the control and manipulation of spin-waves in magnetic materials\,\cite{NG09,KDG10,LUGM11}, the need for theoretical tools capable of accounting for the complex physics of spin-waves is becoming more and more acute. As efficient software packages\,\cite{OOMMF,MuMax,Nmag,MicroMagnum} are now freely available to the scientific community, time-domain numerical micromagnetic simulations\,\cite{KP07}, which are very versatile and allow describing systems of great complexity, tend to become the tool of choice. However, because they are carried out in real space and primarily consist in determining the time evolution of the magnetization spatial distribution, these simulations are not ideally suited to tackle a number of problems of importance in spin-wave studies. One such problem is the determination of the magnetic normal modes, their frequencies and their dispersion relations. Although micromagnetic simulations can be employed with some success for this (see, e.g., Refs.~\onlinecite{MS05,BHF06,YHS07,KLH09,RTB10,GBSA15}), through Fourier analyses of magnetization time series -for determining frequencies- or spatial profiles -for determining wavelengths-, other approaches may be more appropriate. Among the best suited approaches are certainly those based on the dynamic matrix method\,\cite{GGMN04,GMNG04}. They are intrinsically in the frequency domain and, unlike time-domain micromagnetic simulations which can hardly detect low frequency modes and fail altogether to identify degenerate ones, they have the potential to yield all of the normal modes.

The dynamic matrix method is a micromagnetic method, which involves the subdivision of the magnetic medium into small cells, as most micromagnetic numerical schemes, followed by the construction and diagonalization of a matrix that contains all of the information regarding the effective magnetic fields acting on the magnetization vectors of these cells. So far, this method has only been used to study standing spin-waves in confined media such as magnetic nano-elements\,\cite{GGMN04,GMNG04,BHF06,GMFT14} or arrays of dipolar-coupled nanoparticules\,\cite{GMN07}. No scheme has been devised to describe the spin-waves travelling in magnetic media, which are very (infinitely) extended in one or two dimensions. Addressing such spin-waves by means of conventional real-space micromagnetic simulations implies to simulate very long samples (much longer than the longest characteristic wavelength to be calculated). This  requires large amounts of computing power and storage memory, especially when long wavelength spin-waves are considered and a good wave vector resolution is to be achieved. Yet, propagating spin-waves are also essential in magnonic applications, which rely on the capacity to propagate spin excitations along micron-scale circuits in order to process information\,\cite{KBW10}. The main purpose of the present work is thus to deliver the full recipe for a dynamic matrix based numerical scheme adapted to plane (undamped) spin-waves. This is done in the first part of the paper (Secs.~\ref{Sec_Method}-\ref{Sec_dynamic_Demag_Tensor}), where, after describing the principle of the method (Sec.~\ref{Sec_Method}), we derive all the mathematical expressions required for its implementation, i.e., for building the dynamic matrix (Sec.~\ref{Sec_Dynamic_Matrix}), in the case of magnetic media having the shape of extended films or thin strips. As always in micromagnetism, the main difficulty that has to be dealt with lies in the treatment of the long range dipole-dipole interactions. To account for those, we adopt an intuitive approach using demagnetizing tensors. While the concept of demagnetization factors is familiar in the case of homogeneously magnetized cells\,\cite{NWD93}, we show here (Sec.~\ref{Sec_dynamic_Demag_Tensor}) that it can be extended to cells supporting plane spin-waves, that is, cells in which the magnetization vector oscillates harmonically in space. In this respect, our approach can be viewed as the finite-difference counterpart of the magnetostatic Green's function approach used in some (semi-)analytical spin-wave theories\,\cite{KS86,KGHC07}.

The second part of the paper (Sec.~\ref{Sec_Applications}) is devoted to applications of our numerical scheme to questions of current interest. We take this opportunity to illustrate the fact that its most interesting feature is certainly that it allows determining the propagating spin-wave modes in magnetic media which are magnetically inhomogeneous in the plane perpendicular to the direction of propagation, either because the material parameters vary in space or because the equilibrium spin configuration contains a non-collinear magnetic texture. We note that all the numerical data presented in this paper have been obtained with C++ computer programs employing the Eigen template library\,\cite{Eigen} for linear algebra operations, including complex matrix diagonalization, and the GNU Scientific library\,\cite{GSL} for numerical integration.

\section{Principle of the Method\label{Sec_Method}}
In the present work, we are not interested in standing spin-waves inside a ferromagnetic nanoobject, as in the inspiring paper by Grimsditch and coworkers\,\cite{GGMN04}, but rather in spin-waves that propagate as \emph{plane waves}, in a particular direction. This is the reason why we will assume that the magnetic media supporting the spin-waves are unbounded in the direction of propagation, and so will necessarily be the magnetic cells used for discretizing these media. Under such assumptions, only situations where the equilibrium magnetic configuration is invariant upon translation along the propagation direction can be studied. This is the main limitation of our approach.

\begin{figure}[b]
\includegraphics[width=8.5cm,trim=0 66 0 60,clip]{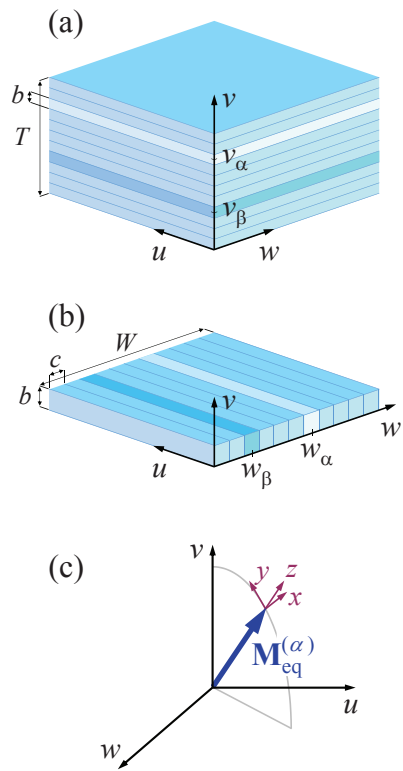}
\caption{(a),(b) Geometries of the magnetic cell arrays. Extended films are subdivided vertically into infinite horizontal slabs (a), whereas strips are subdivided transversally into infinitely long rectangular parallelepipeds (b). (c) $uvw$ and $xyz$ coordinate systems attached to the magnetic medium and equilibrium magnetization, respectively.} \label{Fig_Magnetic_Cells}
\end{figure}

We will consider two types of media, namely films of thickness $T$ [Secs.~\ref{Sec_Films}] and thin strips of width $W$ [Sec.~\ref{Sec_Strips}], which are both compatible with a discretization by means of a one-dimensional array of $N$ geometrically identical cells. For describing our model in more details, we introduce a first coordinate system, $uvw$, and the associated orthonormal direct vector basis $\{\mathbf{e}_u,\mathbf{e}_v,\mathbf{e}_w\}$, such that axis $u$ is parallel to the direction of propagation of the spin-waves and axis $v$ is normal to the film/strip plane. With this, the magnetic cells used to discretize films will be slabs of thickness $b=T/N$, parallel to the $(u,w)$ plane [Fig.~\ref{Fig_Magnetic_Cells}(a)], whereas for strips, the cells will be rectangular parallelepipeds of height $b$ and width $c=W/N$, parallel to axis $u$ [Fig.~\ref{Fig_Magnetic_Cells}(b)].

The starting point of our micromagnetic model is the Landau-Lifshitz equation describing the time evolution of the magnetization vector field $\mathbf{M}(\mathbf{r},t)$ in the absence of magnetic damping, which we linearize around an equilibrium $\mathbf{M}_
{\text{eq}}(\mathbf{r})$. We perform the linearization in the usual way\,\cite{H90}, that is, by writing both $\mathbf{M}(\mathbf{r},t)$ and the effective magnetic field acting on it, $\mathbf{H}_{\text{eff}}(\mathbf{r},t)$, as the sum of a large equilibrium term and a much smaller dynamic term : $\mathbf{M}(\mathbf{r},t) = \mathbf{M}_{\text{eq}}(\mathbf{r})+ \mathbf{m}(\mathbf{r},t)$ with $\mathbf{M}_{\text{eq}}\cdot\mathbf{m}=0$ and $\|\mathbf{M}_{\text{eq}}\|=M_{\text{S}}$, where $M_{\text{S}}$ is the saturation magnetization, and $\mathbf{H}_{\text{eff}}(\mathbf{r},t) =\mathbf{H}_{\text{eq}}(\mathbf{r}) +\mathbf{h}(\mathbf{r},t)$. Keeping only terms up to first order in the small parameters $\mathbf{m}$ and $\mathbf{h}$, the Landau-Lifshitz equation for a particular magnetic cell $\alpha$ becomes
\begin{equation}
\dot{\mathbf{m}}^{(\alpha)}(t) = -|\gamma|\,\mu_0\left[\mathbf{M}_{\text{eq}}^{(\alpha)} \!\times\!\mathbf{h}^{(\alpha)}(t) + \mathbf{m}^{(\alpha)}(t)\!\times\!\mathbf{H}_{\text{eq}}^{(\alpha)}\right]
\label{Eq_LLuvw},
\end{equation}
where $\gamma$ is the gyromagnetic ratio\,\cite{Rem0} and $\mu_0$ is the permeability of free space. Since we are concerned here with plane spin-waves traveling along axis $u$, we postulate that the variable magnetization has the form
\begin{equation}
\mathbf{m}^{(\alpha)}(u,t)=\mathbf{m}_0^{(\alpha)}e^{i(\omega t-ku)},
\label{Eq_PWSW}
\end{equation}
where $\mathbf{m}_0^{(\alpha)}$ is a complex amplitude vector, and $k$ and $\omega$ are the spin-wave wave vector and angular frequency, respectively. Here, it is important to note that $k$ can take on positive and negative values ($\mathbf{k}=k\,\mathbf{e}_u$) in order to account for spin-wave propagation in both $+u$ and $-u$ directions (assuming $\omega\!>\!0$).

In addition to the $uvw$ coordinate system attached to the magnetic medium, we also introduce a coordinate system, $xyz$, and the corresponding vector basis $\{\mathbf{e}_x,\mathbf{e}_y, \mathbf{e}_z\}$, such that axis $z$ is parallel to both $\mathbf{H}_{\text{eq}}^{(\alpha)}$ and $\mathbf{M}_{\text{eq}}^{(\alpha)}$  [Ref.~\,\onlinecite{KS86}], i.e., such that we have $\mathbf{H}_{\text{eq}}^{(\alpha)} = H_{\text{eq}}^{(\alpha)}\,\mathbf{e}_z$ and $\mathbf{M}_{\text{eq}}^{(\alpha)} = M_{\text{S}}\,\mathbf{e}_z$ [Fig.~\ref{Fig_Magnetic_Cells}(c)]. This, together with the fact that Eq.~\ref{Eq_PWSW} implies $\dot{\mathbf{m}}^{(\alpha)}\!=\!i\omega\mathbf{m}^{(\alpha)}$, allows us to reduce the 3-component vector equation for cell $\alpha$ [Eq.~\ref{Eq_LLuvw}] to a set of only two equations
\begin{equation}
\omega \left(\begin{matrix} m_x^{(\alpha)}\\ m_y^{(\alpha)} \end{matrix}\right) = -i|\gamma|\mu_0 \left(\begin{matrix} \;\;\;M_{\text{S}}h_y^{(\alpha)} &-& H_{\text{eq}}^{(\alpha)}\,m_y^{(\alpha)} \\ -M_{\text{S}}h_x^{(\alpha)} &+& H_{\text{eq}}^{(\alpha)}\,m_x^{(\alpha)} \\ \end{matrix}\right),
\label{Eq_LLxyz}
\end{equation}
in accord with the fact that the small amplitude magnetization dynamics around the equilibrium is confined in the $(x,y)$ plane. Let us call $\mathbf{T}^{(\alpha)}$ the transformation matrix from the $uvw$ coordinate system to the \emph{local} coordinate system, $xyz$, attached to $\mathbf{M}_{\text{eq}}^{(\alpha)}$. It is crucial to note that in cases where the equilibrium magnetic configuration is not fully collinear throughout the entire medium (see Sec.~\ref{Sec_Strips}), the $xyz$ coordinate system, hence the matrix $\mathbf{T}^{(\alpha)}$, is not the same for all magnetic cells. Failure to take this into account when necessary inevitably leads to erroneous results.

After discretization, the linearized Landau-Lifshitz equation of the whole magnetic medium thus consists of a set of $2N$ equations, which may finally be written in the form of an eigenvalue equation
\begin{align}
\omega \begin{pmatrix} m_x^{(1)}\\ \vdots \\m_x^{(N)}\\m_y^{(1)}\\ \vdots \\m_y^{(N)} \end{pmatrix} &= \begin{pmatrix}&.\; &\hdots\; &. &.\; &\hdots\; &.& \\ &\vdots\; &\mathbf{D}_{xx}\; &\vdots &\vdots\; &\mathbf{D}_{xy}\; &\vdots& \\ &.\; &\hdots\; &. &.\; &\hdots\; &.& \\&.\; &\hdots\; &. &.\; &\hdots\; &.& \\ &\vdots\; &\mathbf{D}_{yx}\; &\vdots &\vdots\; &\mathbf{D}_{yy}\; &\vdots& \\ &.\; &\hdots\; &. &.\; &\hdots\; &.& \end{pmatrix} \begin{pmatrix} m_x^{(1)}\\ \vdots \\m_x^{(N)}\\m_y^{(1)}\\ \vdots \\m_y^{(N)} \end{pmatrix} \nonumber\\ &= \mathbf{D} \begin{pmatrix} m_x^{(1)}\\ \vdots \\m_x^{(N)}\\m_y^{(1)}\\ \vdots \\m_y^{(N)} \end{pmatrix},
\label{Eq_Eigenvalue}
\end{align}
where $\mathbf{D}$ is the so-called dynamic matrix whose dimensions are $2N\times2N$. How we effectively perform this essential step will be detailed in the next section. If short range exchange and long range dipole-dipole interactions are not included in the model, the $\mathbf{D}$ matrix is block-diagonal\,\cite{GGMN04}. If, on the contrary, those interactions are taken into account, none of the off-diagonal elements is zero and, in general, $\mathbf{D}$ has no special properties. In particular, it is neither hermitian nor (systematically) sparse. Numerical diagonalization of the dynamic matrix is the ultimate step of the process. It yields the profiles of the normal modes across the magnetic medium in the form of ensembles of complex amplitudes $m_{0x}^{(\alpha)}$ and $m_{0y}^{(\alpha)}$ ($\alpha = 1..N$), as well as the corresponding eigenfrequencies, which are real numbers in the absence of magnetic damping.

It is important to note that, irrespective of the sign of $k$, the eigenvectors of $\mathbf{D}$ always come in pairs with eigenfrequencies of (usually) identical absolute values but opposite signs\,\cite{Rem1}, where the eigenvector with $k\omega\!>\!0$ (resp. $k\omega\!<\!0$) corresponds to propagation in the $+u$ direction (resp.~$-u$ direction). Also, for retrieving the true spatiotemporal evolution of the dynamic magnetization in a particular mode, one has to take the real part of $\mathbf{m}^{(\alpha)}$, as defined in Eq.~\ref{Eq_PWSW}. In this operation, all four real and imaginary parts of the $m_{0x}^{(\alpha)}$ and $m_{0y}^{(\alpha)}$ complex amplitudes a priori matter, since $\text{Re}(\mathbf{m}^{(\alpha)})\!= \!\text{Re}(\mathbf{m}_0^{(\alpha)}) \cos(\omega t\!-ku)- \text{Im}(\mathbf{m}_0^{(\alpha)})\sin(\omega t\!-ku)$. In general, $\text{Re}(m_{0x}^{(\alpha)})$, $\text{Im}(m_{0x}^{(\alpha)})$, $\text{Re}(m_{0y}^{(\alpha)})$, and $\text{Im}(m_{0y}^{(\alpha)})$ are all relevant since four parameters are indeed required to fully characterize the precessional motion of the magnetization: three parameters -two radii and a tilt angle $\varphi^{(\alpha)}$- are necessary to describe the elliptical time trajectory of $\tilde{\mathbf{m}}^{(\alpha)}\!= \!\text{Re}(\mathbf{m}^{(\alpha)})$ in the ($x$,$y$) plane, while a fourth one, $\tau^{(\alpha)}$, is needed to account for the relative phase of the precession (see Appendix~\ref{App_Precessional_Motion}). However, situations are rare where the tilt angle $\varphi^{(\alpha)}$ and the phase $\tau^{(\alpha)}$ vary across the profile of a normal mode and thereby cannot be made nil for all magnetic cells\,\cite{Rem2}. This occurs, for instance, when the spin-wave medium is magnetized in-plane and the orientation of the equilibrium magnetization changes in space. Here, no such situation will be considered. Local reference frames and phase origins will be chosen so that the conditions $\varphi^{(\alpha)}\!=\!\tau^{(\alpha)}\!=\!0$ and $\text{Im}(m_{0x}^{(\alpha)})\!=\!\text{Re}(m_{0y}^{(\alpha)})\!=\!0$ are systematically fulfilled and the sole variations of $\text{Re}(m_{0x}^{(\alpha)})$ and $\text{Im}(m_{0y}^{(\alpha)})$ with $\alpha$ suffice to characterize entirely a modal profile.

\section{Construction of the dynamic matrix\label{Sec_Dynamic_Matrix}}
Examination of equation~\ref{Eq_LLxyz} shows that the construction of the dynamic matrix requires essentially two things. The first one is to evaluate the magnitude of the static part of the effective magnetic field in each cell, $H_{\text{eq}}^{(\alpha)}$, knowing what the whole equilibrium magnetic configuration $\mathbf{M}_{\text{eq}}^{(\beta)}$ (with $\beta = 1..N$) is. This is rather trivial. As in usual micromagnetic simulations, one needs to take into account contributions from the external magnetic field ($\mathbf{H}_0$), crystal anisotropy ($\mathbf{H}_{\text{K}}$), exchange interactions between nearest neighbor cells ($\mathbf{H}_{\text{ex}}$), and dipolar interactions ($\mathbf{H}_{\text{d}}$). All of these are more easily evaluated in the $uvw$ coordinate system.

For crystal anisotropy, we use the general expression
\begin{align}
\mathbf{H}_{\text{K}}^{(\alpha)} =& \frac{2K_{\text{u}}}{\mu_0M_{\text{S}}^2} \left(\mathbf{M}_{\text{eq}}^{(\alpha)}\cdot\mathbf{a} \right)\mathbf{a} \nonumber\\ &- \frac{2K_{\text{c}}}{\mu_0M_{\text{S}}^4}\sum_{i=1}^{3}\left[ \sum_{j\neq i}(\mathbf{M}_{\text{eq}}^{(\alpha)}\cdot \mathbf{c}_j)^2\right] \!\left(\mathbf{M}_{\text{eq}}^{(\alpha)}\cdot \mathbf{c}_i\right) \mathbf{c}_i.
\label{Eq_Ha}
\end{align}
The first term accounts for a uniaxial anisotropy of constant $K_\text{u}$ and axis $\mathbf{a}$, while the second stands for a cubic anisotropy of constant $K_\text{c}$ and axes $\mathbf{c}_1$, $\mathbf{c}_2$, and $\mathbf{c}_3$ such that $\mathbf{c}_i\cdot\mathbf{c}_j = \delta_{ij}$, where $\delta_{ij}$ is Kronecker's delta. Both terms are first-order.

In the continuous-medium approximation, the usual expression for isotropic exchange is \begin{equation}
\mathbf{H}_{\text{ex}}(\mathbf{r}) = \Lambda^2\,\Delta\mathbf{M}_{\text{eq}}(\mathbf{r}),
\label{Eq_Hex1}
\end{equation}
where $\Delta$ is the Laplacian operator and $\Lambda\!= \!\sqrt{\frac{2A}{\mu_0M_{\text{S}}^2}}$ is the exchange length, with $A$ the exchange stiffness constant. Using a discrete expression of the second central derivative based on second-order Taylor expansion and implementing free boundary conditions in the simplest possible way (see. Sec.~\ref{Sec_Extensions}), it becomes
\begin{align}
\mathbf{H}_{\text{ex}}^{(\alpha)}=\frac{\Lambda^2}{\xi^2}&\left[\left(
\mathbf{M}_{\text{eq}}^{(\alpha\!-\!1)} - \mathbf{M}_{\text{eq}}^{(\alpha)}\right)(1\!-\!\delta_{1\alpha}) \right. \nonumber\\ &- \left.\left(\mathbf{M}_{\text{eq}}^{(\alpha)} - \mathbf{M}_{\text{eq}}^{(\alpha\!+\!1)}\right)(1\!-\!\delta_{N\alpha})\right]
\label{Eq_Hex2}
\end{align}
where $\xi\!=\!b$ (films) or $c$ (strips), depending on the geometry of the magnetic cell array [Fig.~\ref{Fig_Magnetic_Cells}(a,b)].

Finally, for dipole-dipole interactions, we follow a conventional micromagnetic approach, where the dipolar field experienced by the magnetization of cell $\alpha$ is related to the magnetization vectors of cells $\beta=1..N$, which create the field, through demagnetizing tensors\,\cite{NWD93}
\begin{equation}
\mathbf{H}_{\text{d}}^{(\alpha)} = -\sum_{\beta=1}^{N}\mathbf{N}^{(\alpha\beta)}\cdot \mathbf{M}_{\text{eq}}^{(\beta)}.
\label{Eq_Hd}
\end{equation}

The dimensionless tensor $\mathbf{N}^{(\alpha\beta)}$ is necessarily symmetric and it depends on the shape and relative position of the source $(\beta)$ and target $(\alpha)$ cells. In the case of infinitely extended slabs [Fig.~\ref{Fig_Magnetic_Cells}(a)], mutual $(\alpha\!\neq\!\beta)$ demagnetizing effects are strictly nil and only the $v$-component of the magnetization creates a self $(\alpha\!=\!\beta)$ demagnetizing field, which is along $v$. Thus, in the $uvw$ coordinate system, $\mathbf{N}^{(\alpha\beta)}$ takes the trivial form
\begin{equation}
\mathbf{N}^{(\alpha\beta)}=
\begin{vmatrix}
\,&0& &0& &0&\\
\,&0& &\delta_{\alpha\beta}& &0&\\
\,&0&& 0& &0&\\
\end{vmatrix}.
\label{Eq_Slab_N}
\end{equation}

In the case of infinitely long cells with rectangular $(b\!\times\!c)$ cross section, general analytical expressions can be derived for the non-zero components of $\mathbf{N}^{(\alpha\beta)}$, which are quite cumbersome (see Appendix~\ref{App_Static_Demag_Tensor}). Under the assumption that the cells are arranged in a one-dimensional array [Fig.~\ref{Fig_Magnetic_Cells}(b)], they become somewhat simpler and the demagnetizing tensor reduces to
\begin{equation}
\mathbf{N}^{(\alpha\beta)}=
\begin{vmatrix}
\;\,0& 0& 0&\\
\;\,0& N_{vv}^{(\alpha\beta)}& 0&\\
\;\,0& 0& N_{ww}^{(\alpha\beta)}&
\end{vmatrix}.
\label{Eq_Rect_N}
\end{equation}
Moreover, as a consequence of $\text{tr}(\mathbf{N}^{(\alpha\beta)})=\delta_{\alpha\beta}$ [Ref.~\onlinecite{NWD93}], the two non-zero diagonal components are conveniently related to each other by
\begin{equation}
N_{vv}^{(\alpha\beta)}+N_{ww}^{(\alpha\beta)} = \delta_{\alpha\beta}.
\label{Eq_Trace_N}
\end{equation}
Thus, calculating $\mathbf{N}^{(\alpha\beta)}$ simply amounts to evaluating a single quantity, for instance $N_{ww}^{(\alpha\beta)}$. When the source and target cells coincide (self demagnetizing tensor), this tensor element depends solely on the aspect ratio $p=c/b$. It is given by the equation
\begin{equation}
N_{ww}(p) = \frac{1}{\pi}\left[\frac{1-p^2}{2p}\ln\left(1+p^2\right)+p\ln\!p + 2\arctan\!\left(\frac{1}{p}\right)\right]
\label{Eq_Rect_Nww_Self}
\end{equation}
derived by Brown\,\cite{B62} and Aharoni\,\cite{A98}. When, on the contrary, the source cell is located at a finite distance $n_cc$ (with $n_c\in\mathbb{N}^*$) from the target cell (mutual demagnetizing tensor), it becomes
\begin{widetext}
\begin{equation}
N_{ww}^{(\alpha\beta)}(p,n_c) = \sum_{n=n_c-1}^{n_c+1} \frac{(-2)^{1-|n-n_c|} }{2\pi} \left[\frac{1-n^2p^2}{2p}\ln\!\left(1+n^2p^2\right) + n^2p\ln\!\left|np\right| + 2n\arctan\!\left(\frac{1}{np}\right)\right],
\label{Eq_Rect_Nww_Mut}
\end{equation}
\end{widetext}
which follows from Eq.~\ref{Eq_Static_Nww} in the particular case $\delta v=0$, $\delta w=n_c\,c$.

\vspace{0.4cm}
The second task we need to perform for building $\mathbf{D}$ is to express the $x$ and $y$ components of the dynamic field in each cell  $\{h_x^{(\alpha)}$, $h_y^{(\alpha)}\}$ in terms of the $x$ and $y$ components of the variable magnetization in every cells  $\{m_x^{(1)},..,m_x^{(N)},m_y^{(1)},.., m_y^{(N)}\}$, in an explicit manner. This is a slightly more complex task, especially because dipole-dipole interactions couple all the individual Landau-Lifshitz equations [Eq.~\ref{Eq_LLxyz}] together. Special attention must also be paid since we now have to express all vector quantities in the more relevant $xyz$ coordinate system.

As for the equilibrium magnetic field, we must add up several contributions to obtain the full dynamic magnetic field $\mathbf{h}^{(\alpha)}$ produced when the magnetization departs from equilibrium ($\mathbf{m}^{(\beta)}\neq\mathbf{0}$). With the exception of the external field $\mathbf{H}_0$, which is assumed to be time-independent, all contributions to $\mathbf{H}_{\text{eq}}^{(\alpha)}$ have a dynamic counterpart. The contribution of crystal anisotropy to $\mathbf{h}^{(\alpha)}$, which corresponds to the equilibrium anisotropy field in Eq.~\ref{Eq_Ha}, is\,\cite{Rem3}
\begin{widetext}
\begin{align}
\mathbf{h}_{\text{K}}^{(\alpha)}
=&\frac{2K_{\text{u}}}{\mu_0M_{\text{S}}^2}\left(\mathbf{m}^{(\alpha)}\cdot \mathbf{a}\right)\,\mathbf{a}
\nonumber\\
&-\frac{2K_{\text{c}}}{\mu_0M_{\text{S}}^4}\sum_{i=1}^{3}\left\{\left[ \sum_{j\neq i}\left(\mathbf{M}_{\text{eq}}^{(\alpha)}\!\cdot\! \mathbf{c}_j\right)^2\right]\!\left(\mathbf{m}^{(\alpha)}\!\cdot\! \mathbf{c}_i\right)\mathbf{c}_i + 2\!\left[\sum_{j\neq i}
\left(\mathbf{M}_{\text{eq}}^{(\alpha)}\!\cdot\!\mathbf{c}_j\right) \left(\mathbf{m}^{(\alpha)}\!\cdot\!\mathbf{c}_j\right) \right] \!\left(\mathbf{M}_{\text{eq}}^{(\alpha)}\!\cdot\!\mathbf{c}_i\right) \mathbf{c}_i \right\},
\label{Eq_ha1}
\end{align}
in a compact vector form. Getting the $x$ and $y$ components of $\mathbf{h}_{\text{K}}^{(\alpha)}$ as a function of those of $\mathbf{m}^{(\alpha)}$ from Eq.~\ref{Eq_ha1} is a matter of simple arithmetics. We obtain
\begin{align}\label{Eq_ha2}
(l=x,y)\qquad
\mathbf{e}_l\cdot\mathbf{h}_{\text{K}}^{(\alpha)} =& \frac{2K_{\text{u}}} {\mu_0M_{\text{S}}^2}\left[ a_la_x\;m_x^{(\alpha)} + a_la_y\;m_y^{(\alpha)}\right]
\nonumber\\
&-\frac{2K_{\text{c}}}{\mu_0M_{\text{S}}^2} \left\{
\vphantom{\left.\right]m_y^{(\alpha)}}
\left[\; c_{1l}c_{1x}\left(c_{2z}^2+c_{3z}^2\right) +2c_{2z}c_{3z}\left(c_{2x}c_{3l}+c_{2l}c_{3x}\right)
\right.\right. \nonumber\\ &\qquad\qquad\;\;\,
\vphantom{\left.\right]m_y^{(\alpha)}}
+\!c_{2l}c_{2x}\left(c_{3z}^2+c_{1z}^2\right) +2c_{3z}c_{1z}\left(c_{3x}c_{1l}+c_{3l}c_{1x}\right)
\nonumber\\ &\left.\qquad\qquad\;\,
\vphantom{\left.\right]m_y^{(\alpha)}}
+\!c_{3l}c_{3x}\left(c_{1z}^2+c_{2z}^2\right) +2c_{1z}c_{2z}\left(c_{1x}c_{2l}+c_{1l}c_{2x}\right)\right] m_x^{(\alpha)}
\nonumber\\ &\qquad\quad\;\;\;
\vphantom{\left.\right]m_y^{(\alpha)}}
+\left[ \;c_{1l}c_{1y}\left(c_{2z}^2+c_{3z}^2\right) +2c_{2z}c_{3z}\left(c_{2y}c_{3l}+c_{2l}c_{3y}\right)
\right. \nonumber\\
&\qquad\qquad\;\;\,
\vphantom{\left.\right]m_y^{(\alpha)}}
+\!c_{2l}c_{2y}\left(c_{3z}^2+c_{1z}^2\right) +2c_{3z}c_{1z}\left(c_{3y}c_{1l}+c_{3l}c_{1y}\right)
\nonumber\\ &\left.\left.\qquad\qquad\;\;
+c_{3l}c_{3y}\left(c_{1z}^2+c_{2z}^2\right) +2c_{1z}c_{2z}\left(c_{1y}c_{2l}+c_{1l}c_{2y}\right)\right] m_y^{(\alpha)}\right\},
\end{align}
\end{widetext}
where $(a_x,a_y,a_z)$ and $(c_{ix},c_{iy},c_{iz})$ are the coordinates of the unit vectors $\mathbf{a}$ and $\mathbf{c}_i$ in the $\{\mathbf{e}_x, \mathbf{e}_y,\mathbf{e}_z\}$ local basis. By definition of the transformation matrix $\mathbf{T}^{(\alpha)}$, those are related to the coordinates in the $\{\mathbf{e}_u, \mathbf{e}_v,\mathbf{e}_w\}$ global basis by $v_l=\sum_{j=1}^{3}T^{(\alpha)}_{lj}v_j$ $(\mathbf{v}=\mathbf{a}, \mathbf{c}_i)$, where the indices $l,j=1,2,3$ stand for $x,y,z$ and $u,v,w$, respectively.

The contribution of isotropic exchange to the dynamic magnetic field,  $\mathbf{h}_{\text{ex}}(\mathbf{r})$, in the continuous-medium approximation, may be obtained by replacing $\mathbf{M}_{\text{eq}}(\mathbf{r})$ with $\mathbf{m}(\mathbf{r})$ in Eq.~\ref{Eq_Hex1}. Applied to plane spin-waves of the form given in Eq.~\ref{Eq_PWSW} and introducing free boundary conditions, the expression one obtains becomes
\begin{widetext}
\begin{equation}
\mathbf{h}_{\text{ex}}^{(\alpha)} = \frac{\Lambda^2}{\xi^2}\left[\left( \mathbf{m}^{(\alpha\!-\!1)}\!-\!\mathbf{m}^{(\alpha)}\right) (1\!-\!\delta_{1\alpha}) - \left(\mathbf{m}^{(\alpha)}\!-\!\mathbf{m}^{(\alpha\!+\!1)}\right) (1\!-\!\delta_{N\alpha})\right]-\Lambda^2k^2\mathbf{m}^{(\alpha)}
\label{Eq_hex1}
\end{equation}
\end{widetext}
after discretization. Deriving correct expressions for the $x$ and $y$ components of $\mathbf{h}_{\text{ex}}^{(\alpha)}$ as a function of those of the variable magnetization vectors $\mathbf{m}^{(\alpha\!-\!1)}$, $\mathbf{m}^{(\alpha)}$, and $\mathbf{m}^{(\alpha\!+\!1)}$ from Eq.~\ref{Eq_hex1} requires to take into account the possible non-collinear character of the equilibrium, i.e., the fact that the local $xyz$ reference frames in the nearest neighbor cells $\alpha\pm\!1$ are not necessarily the same as in cell $\alpha$. This can be achieved by introducing the relevant transformation matrices, $\mathbf{T}^{(\beta)}$, with $\beta=\alpha\!-\!1,\alpha,\alpha\!+\!1$, and their inverse matrices $\bar{\mathbf{T}}^{(\beta)}=\left(\mathbf{T}^{(\beta)}\right)^{-1}$. With $T^{(\beta)}_{ij}$ (resp. $\bar{T}^{(\beta)}_{ij}$) denoting the elements of matrix $\mathbf{T}^{(\beta)}$ (resp. $\bar{\mathbf{T}}^{(\beta)}$), we have
\begin{widetext}
\begin{align}\label{Eq_hex2}
(l=x,y)\qquad
\mathbf{e}_l\cdot\mathbf{h}_{\text{ex}}^{(\alpha)} = \frac{\Lambda^2}{\xi^2}
&\left[\;\left(\sum_{i=1}^{3}T^{(\alpha)}_{li}
\bar{T}^{(\alpha\!-\!1)}_{i1}\right)(1\!-\!\delta_{1\alpha})
\;m_x^{(\alpha\!-\!1)}
+\left(\sum_{i=1}^{3}T^{(\alpha)}_{li}
\bar{T}^{(\alpha\!-\!1)}_{i2}\right)(1\!-\!\delta_{1\alpha})
\;m_y^{(\alpha\!-\!1)}\right.
\nonumber\\
&-\left(\sum_{i=1}^{3}T^{(\alpha)}_{li}\bar{T}^{(\alpha)}_{i1}\right)
\left(2+\xi^2k^2\right)m_x^{(\alpha)}
-\left(\sum_{i=1}^{3}T^{(\alpha)}_{li}\bar{T}^{(\alpha)}_{i2}\right)
\left(2+\xi^2k^2\right)m_y^{(\alpha)}
\nonumber\\
&\left.+\left(\sum_{i=1}^{3}T^{(\alpha)}_{li}
\bar{T}^{(\alpha\!+\!1)}_{i1}\right)(1\!-\!\delta_{N\alpha})
\;m_x^{(\alpha\!+\!1)}
+\left(\sum_{i=1}^{3}T^{(\alpha)}_{li}
\bar{T}^{(\alpha\!+\!1)}_{i2}\right)(1\!-\!\delta_{N\alpha})
\;m_y^{(\alpha\!+\!1)}\right],
\end{align}
\end{widetext}
where the indices $l\!=\!x$ and $l\!=\!y$ are to be understood as $l\!=\!1$ and $l\!=\!2$, respectively, when it comes to the elements of matrix $\mathbf{T}^{(\alpha)}$.

Finally, for describing dynamic dipole-dipole interactions between magnetic cells, we use also demagnetizing tensors, as in the static case. By analogy with Eq.~\ref{Eq_Hd}, we write the dipolar contribution to $\mathbf{h}^{(\alpha)}$ as
\begin{equation}
\mathbf{h}_{\text{d}}^{(\alpha)} = -\sum_{\beta=1}^{N}\mathbf{n}^{(\alpha\beta)}\cdot \mathbf{m}^{(\beta)}.
\label{Eq_hd1}
\end{equation}
We will see in the next section that the components of the newly introduced \emph{plane-wave demagnetizing tensor} $\mathbf{n}^{(\alpha\beta)}$ depend not only on the geometry and relative positions of the magnetic cells, as usual (static) demagnetization factors do, but also on the wave vector $k$ of the propagating spin-waves. We will also see that rather simple analytical expressions can be derived for $\mathbf{n}^{(\alpha\beta)}$ when the cells are extended slabs [Fig.~\ref{Fig_Magnetic_Cells}(a)] but that complex integral expressions must be dealt with in case the cells are rectangular parallelepipeds [Fig.~\ref{Fig_Magnetic_Cells}(b)]. When deriving expressions for the $x$ and $y$ components of $\mathbf{h}_{\text{d}}^{(\alpha)}$ as a function of those of the variable magnetization vectors $\mathbf{m}^{(\beta)}$ ($\beta=1..N$) from Eq.~\ref{Eq_hd1}, we must once again account for the possible non-collinear character of the equilibrium by introducing the appropriate transformation matrices. With tensors $\mathbf{n}^{(\alpha\beta)}$ expressed in the $uvw$ coordinate system, we have $\mathbf{h}_{\text{d}}^{(\alpha)} = -\sum_{\beta=1}^{N} \mathbf{T}^{(\alpha)} \mathbf{n}^{(\alpha\beta)} \bar{\mathbf{T}}^{(\beta)} \cdot \mathbf{m}^{(\beta)}$ in the $xyz$ reference frame. It readily follows
\begin{widetext}
\begin{align}\label{Eq_hd2}
(l=x,y)\qquad
\mathbf{e}_l\cdot\mathbf{h}_{\text{d}}^{(\alpha)} &= -\sum_{\beta=1}^{N}
\left\{\left[\sum_{i=1}^{3} \left(\sum_{j=1}^{3} T^{(\alpha)}_{lj} n^{(\alpha\beta)}_{ji}\right) \bar{T}^{(\beta)}_{i1}\right]m_x^{(\beta)}
+\!
\left[\sum_{i=1}^{3} \left(\sum_{j=1}^{3} T^{(\alpha)}_{lj} n^{(\alpha\beta)}_{ji}\right) \bar{T}^{(\beta)}_{i2}\right]m_y^{(\beta)}
\right\},
\end{align}
\end{widetext}
where indices $i,j=1,2,3$ stand for $u,v,w$ when it comes to the elements of tensor $\mathbf{n}^{(\alpha\beta)}$ and, as before, the indices $l\!=\!x$ and $l\!=\!y$ are to be understood as $l\!=\!1$ and $l\!=\!2$ when it comes to the elements of matrix $\mathbf{T}^{(\alpha)}$.

For the sake of simplicity, we have implicitly assumed so far that the magnetic parameters of the media supporting the spin-waves were homogeneous. While accounting for space variations of magnetic anisotropy is straightforward (anisotropy constants and axes just need to be made $\alpha$-dependent in Eqs.~\ref{Eq_Ha}, \ref{Eq_ha1}, and \ref{Eq_ha2}), introducing space variations of saturation magnetization and/or exchange stiffness is not. Indeed, to allow for an $\alpha$-dependence of these two parameters, the expressions of the static [Eq.~\ref{Eq_Hex2}] and dynamic [Eq.~\ref{Eq_hex1}] exchange fields must be transformed in a way which is not totally trivial, as we will discuss in Sec.~\ref{Sec_Films}.

\section{Dynamic demagnetizing tensors\label{Sec_dynamic_Demag_Tensor}}
Our goal in this key section is to derive mathematical expressions for the non-zero components of the dynamic demagnetizing tensors $\mathbf{n}^{(\alpha\beta)}$ introduced earlier [Sec.~\ref{Sec_Dynamic_Matrix}, Eq.~\ref{Eq_hd1}]. Since these expressions are among the most important results of the present work, we shall give a rather detailed description of how they can be obtained. In short, what needs to be done is the following. First, the nature of the magnetic charges that the variable magnetization of the source cell creates must be determined. Next, the magnetic field that these charges produce must be evaluated and averaged throughout the target cell. Last, the tensor element $n_{ij}^{(\alpha\beta)}$ must be identified with the negative of the proportionality factor between the $i$-component of the averaged field and the $j$-component of the variable magnetization.

At this point, it is important to note that $\mathbf{n}^{(\alpha\beta)}$ has the same intrinsic properties as all demagnetizing tensors\,\cite{NWD93}: It is symmetric and obeys $\text{tr}(\mathbf{n}^{(\alpha\beta)}) = \delta_{\alpha\beta}$. Furthermore, as its static counterpart, $\mathbf{n}^{(\alpha\beta)}$ is necessarily diagonal when the source and target cells coincide $(\alpha\!=\!\beta)$. All these requirements reduce very strongly the number of independent tensor elements that need to be calculated to fully determine $\mathbf{n}^{(\alpha\beta)}$, in general.

\subsection{Infinite slabs\label{Sec_Slabs}}
In the case where the magnetic cells are slabs with infinite dimensions along both $u$ and $w$, the in-plane component of the variable magnetization perpendicular to the direction of propagation, i.e., the $w$-component, never produces any magnetic charge. Therefore, the corresponding tensor elements $n_{iw}^{(\alpha\beta)}\!=\!n_{wi}^{(\alpha\beta)}$ (with $i=u,v,w$) are always nil and $\mathbf{n}^{(\alpha\beta)}$ contains at most four non-zero components. Volume charges $-\boldsymbol\nabla\!\cdot\!\mathbf{m}^{(\beta)}\!= \! i\,k\,m_{u}^{(\beta)}$ are created by the $u$-component of $\mathbf{m}^{(\beta)}$, as a result of its space oscillatory nature [Eq.~\ref{Eq_PWSW}], and surface charges exist at the top $(+m_{v}^{(\beta)})$ and bottom $(-m_{v}^{(\beta)})$ surfaces of the source cell as soon as $\mathbf{m}^{(\beta)}$ has a finite vertical component. With $\mathbf{m}^{(\beta)}$ having the form of a plane wave, all those magnetic charges vary harmonically along the direction of propagation.

Let us then consider a magnetic surface charge distribution parallel to the $(u,w)$ plane, located at the vertical position $v_0$, and harmonic in the $u$-direction, as defined by
\begin{equation}
\sigma_{\text{2D}}(v_0,\mathbf{r},t) = \sigma_0\,\delta(v\!-\!v_0) \,e^{i(\omega t-ku)},
\label{Eq_2D_Charge_Dist}
\end{equation}
where $\sigma_0$ is a complex amplitude and $\delta$ is Dirac's function. The magnetostatic potential created by such a charge distribution writes\,\cite{B11}
\begin{equation}
\phi_{\text{2D}}(\sigma_0,v_0,\mathbf{r},t) =
\frac{\sigma_0}{2|k|}\,e^{-|k(v-v_0)|}\,e^{i(\omega t-ku)},
\label{Eq_Slab_Magn_Potential}
\end{equation}
and the magnetic field that derives from it is
\begin{align}
\mathbf{h}_{\text{2D}}(\sigma_0,v_0,\mathbf{r},t) =& -\!\boldsymbol\nabla\phi_{\text{2D}}(\sigma_0,v_0,\mathbf{r},t)\nonumber\\
=&\frac{\sigma_0}{2}\,e^{-|k(v-v_0)|}\,e^{i(\omega t-ku)}\nonumber\\
&\times \left[i\,\text{sgn}(k)\,\mathbf{e}_u+\text{sgn}(v\!-\!v_0)\,\mathbf{e}_v\right].
\label{Eq_Slab_Magn_Field}
\end{align}
With Eq.~\ref{Eq_Slab_Magn_Field}, we can evaluate any component of $\mathbf{n}^{(\alpha\beta)}$. We shall illustrate this by calculating two particular dynamic demagnetization factors, $n_{uu}^{(\alpha\beta)}$ and $n_{uv}^{(\alpha\beta)}$, which is a priori enough to fully determine the tensor in case the magnetic cells have the shape of infinitely extended slabs.

Let $v_{\beta}$ and $v_{\alpha}$ be the vertical coordinates of the source and target cells, respectively. To derive the expression of $n_{uu}^{(\alpha\beta)}$, we need to calculate the $u$-component of the magnetic field produced by the volume charges $ikm_{u}^{(\beta)}$ in the source cell and then average it over the thickness of the target cell. The first step amounts to integrating the contributions of all surface charge distributions $ikm_{u}^{(\beta)}\delta(v\!-\!v_0)= ikm_{0u}^{(\beta)}\,\delta(v\!-\!v_0)\,e^{i(\omega t-ku)}$ such that $v_{\beta}\!-\!b/2\!\leqslant\!v_0\!\leqslant\!v_{\beta}\!+\!b/2$.
Therefore, we can write
\begin{equation}
-n_{uu}^{(\alpha\beta)}m_{u}^{(\beta)} = \frac{1}{b}
\int\limits_{v_{\alpha}-\frac{b}{2}}^{v_{\alpha}+\frac{b}{2}}\!
\int\limits_{v_{\beta}-\frac{b}{2}}^{v_{\beta}+\frac{b}{2}}\!
\mathbf{e}_u\cdot\!\mathbf{h}_{\text{2D}}\!\left(\!ikm_{0u}^{(\beta)},v_0, \mathbf{r},t\right)\!dv_0\,dv.
\label{Eq_Slab_nuu1}
\end{equation}
Substituting Eq.~\ref{Eq_Slab_Magn_Field} in Eq.~\ref{Eq_Slab_nuu1}, it comes
\begin{equation}
n_{uu}^{(\alpha\beta)} = \frac{|k|}{2b}
\int_{v_{\alpha}-\frac{b}{2}}^{v_{\alpha}+\frac{b}{2}}\!
\int_{v_{\beta}-\frac{b}{2}}^{v_{\beta}+\frac{b}{2}}\!
e^{-|k(v-v_0)|}\,dv_0\,dv.
\label{Eq_Slab_nuu2}
\end{equation}
The result of this double integration, which can be carried out analytically, is different depending on whether $\alpha\!=\!\beta$, or not. If the source and target cells coincide, we obtain
\begin{equation}
(\alpha\!=\!\beta)\qquad n_{uu} = 1 - \frac{1-e^{-|k|b}}{|k|b},
\label{Eq_Slab_nuu_Self}
\end{equation}
whereas, if they are disjoint, we have
\begin{equation}
(\alpha\!\neq\!\beta)\qquad n_{uu}^{(\alpha\beta)} =
\frac{2\,\sinh^2\left(\frac{kb}{2}\right)}{|k|b} \,e^{-|k(v_{\alpha}-v_{\beta})|}.
\label{Eq_Slab_nuu_Mutual}
\end{equation}
As we could have anticipated, when $\alpha\!=\!\beta$, $n_{uu}$ corresponds to the well-known $P_{00}$ coefficient, which appears in the expression of the matrix elements of the dipole-dipole interaction in the popular spin-waves theory of Kalinikos and Slavin\,\cite{KS86,K81}, as the result of the integration over the film(cell) thickness of the magnetostatic Green's function for a plane spin-wave having a uniform profile.

To calculate $n_{uv}^{(\alpha\beta)}$, we have to consider this time the $u$-components of the magnetic fields produced by the two surface charge distributions $\pm\,m_{0v}^{(\beta)}\,\delta(v\!-v_{\beta}\!\mp\!b/2) \,e^{i(\omega t-ku)}$, add them up, and average the sum over the target cell. Therefore, we can write
\begin{align}
-n_{uv}^{(\alpha\beta)}m_{v}^{(\beta)} = \frac{1}{b}\int\limits_{v_{\alpha}-\frac{b}{2}}^{v_{\alpha}+\frac{b}{2}}\!  \mathbf{e}_u\cdot\! & \left[\mathbf{h}_{\text{2D}}\! \left(\!+m_{0v}^{(\beta)},v_{\beta}\!+\!\frac{b}{2}, \mathbf{r},t\right)\right. \nonumber\\ &\left. +\mathbf{h}_{\text{2D}}\!\left(\!-m_{0v}^{(\beta)}, v_{\beta}\!-\!\frac{b}{2},\mathbf{r},t\right)
\right]\!dv.
\label{Eq_Slab_nuv1}
\end{align}
Substituting Eq.~\ref{Eq_Slab_Magn_Field} in Eq.~\ref{Eq_Slab_nuv1}, it comes
\begin{align}
n_{uv}^{(\alpha\beta)} =& -\frac{i}{2b}\,\text{sgn}(k) \nonumber\\
&\times \int_{v_{\alpha}-\frac{b}{2}}^{v_{\alpha}+\frac{b}{2}}\!\left[
e^{-\left|k\left(v-v_{\beta}-\frac{b}{2}\right)\!\right|} -
e^{-\left|k\left(v-v_{\beta}+\frac{b}{2}\right)\!\right|}\right]dv.
\label{Eq_Slab_nuv2}
\end{align}
As for $n_{uu}^{(\alpha\beta)}$, the integration can be performed analytically and we find that the result is different depending on whether the source and target cells coincide or not. When they do coincide, $n_{uv}^{(\alpha\beta)}$ is zero, which means that the dynamic self demagnetizing tensor is diagonal, as expected. It can be written as
\begin{equation}
(\alpha\!=\!\beta)\qquad \mathbf{n}^{(\text{self})}=
\begin{vmatrix}
\,&n_{uu}& &0& &0&\\
\,&0& &1-n_{uu}& &0&\\
\,&0&& 0& &0&\\
\end{vmatrix},
\label{Eq_Slab_n_Self}
\end{equation}
with $n_{uu}$ given by Eq.~\ref{Eq_Slab_nuu_Self}. When the source and target cells are disjoint, we obtain
\begin{equation}
n_{uv}^{(\alpha\beta)} = -i\,\text{sgn}(k)\,\text{sgn}(v_{\alpha}\!-\!v_{\beta})\, \frac{2\sinh^2\left(\frac{kb}{2}\right)}{|k|b}\, e^{-\left|k\left(v_{\alpha}-v_{\beta}\right)\right|}.
\label{Eq_Slab_nuv_Mutual}
\end{equation}
Comparing Eqs.~\ref{Eq_Slab_nuu_Mutual} and \ref{Eq_Slab_nuv_Mutual}, it appears that $n_{uu}^{(\alpha\beta)}$ and $n_{uv}^{(\alpha\beta)}$ are related to each other by
\begin{equation}
n_{uv}^{(\alpha\beta)} = -i\,\text{sgn}(k)\,\text{sgn}(v_{\alpha}\!- \!v_{\beta})\;n_{uu}^{(\alpha\beta)}.
\label{Eq_Slab_nuv_Mutual_bis}
\end{equation}
The dynamic mutual demagnetizing tensor can then be written as
\begin{widetext}
\begin{equation}
(\alpha\!\neq\!\beta)\;\;\;\; \mathbf{n}^{(\alpha\beta)} =
\begin{vmatrix}
\,&n_{uu}^{(\alpha\beta)}& &-i\,\text{sgn}(k)\,\text{sgn}(v_{\alpha}\!- \!v_{\beta})\,n_{uu}^{(\alpha\beta)}& &0&\\
\,&-i\,\text{sgn}(k)\,\text{sgn}(v_{\alpha}\!- \!v_{\beta})\,n_{uu}^{(\alpha\beta)}& &-n_{uu}^{(\alpha\beta)}& &0&\\
\,&0&& 0& &0&\\
\end{vmatrix},
\label{Eq_Slab_n_Mutual}
\end{equation}
\end{widetext}
with $n_{uu}^{(\alpha\beta)}$ given by Eq.~\ref{Eq_Slab_nuu_Mutual}. Thus, for magnetic cells in the shape of extended slabs, a single demagnetization factor is always sufficient to fully determine the whole tensor $\mathbf{n}^{(\alpha\beta)}$.

\subsection{Rectangular parallelepipeds\label{Sec_Parallelepipeds}}
In the case where the magnetic cells are rectangular parallelepipeds with infinite length along $u$, the situation is more complex, both physically and mathematically. In addition to the volumes charges $-\boldsymbol\nabla\!\cdot\!\mathbf{m}^{(\beta)}\!=\! i\,k\,m_{u}^{(\beta)}$ and surface charges on the top $(+m_{v}^{(\beta)})$ and bottom $(-m_{v}^{(\beta)})$ faces of the source cell created, as before, by the $u$ and $v$ components of $\mathbf{m}^{(\beta)}$, surface charges are also generated now on the left $(+m_{w}^{(\beta)})$ and right $(-m_{w}^{(\beta)})$ faces of the cell by the $w$-component of $\mathbf{m}^{(\beta)}$. Therefore, none of the elements of $\mathbf{n}^{(\alpha\beta)}$ is a priori zero, at least when the source and target cells are disjoint.

For calculating the dynamic demagnetization factors in the same manner as above [Sec.~\ref{Sec_Slabs}], we now need to know the expression of the magnetic field, $\mathbf{h}_{\text{1D}}$, emanating from a one-dimensional harmonic distribution of magnetic charges parallel to axis $u$ and located at the transverse position $(v_0,w_0)$, as defined by
\begin{equation}
\sigma_{\text{1D}}(v_0,w_0,\mathbf{r},t) = \sigma_0 \,\delta(v\!-\!v_0) \,\delta(w\!-\!w_0) \,e^{i(\omega t-ku)},
\label{Eq_1D_Charge_Dist}
\end{equation}
where $\sigma_0$ is again a complex amplitude. To our knowledge, no such expression is available in the literature. It is derived from basic magnetostatics in Appendix~\ref{App_Magn_Field_From Harmonic_Dist}. In the $uvw$ reference frame, the three components of $\mathbf{h}_{\text{1D}}$ are
\begin{widetext}
\begin{subequations}\label{Eq_Rect_Magn_Field}
\begin{align}
&\mathbf{e}_u\cdot\mathbf{h}_{\text{1D}}(\sigma_0,v_0,w_0,\mathbf{r},t)\,= i\,\frac{\sigma_0}{2\pi}\;\,k\;\, K_0(|k|\sqrt{(v\!-\!v_0)^2\!+\!(w\!-\!w_0)^2}\,) \,e^{i(\omega t-ku)}
\label{Eq_u_comp_Rect_Magn_Field}\\
&\mathbf{e}_v\cdot\mathbf{h}_{\text{1D}}(\sigma_0,v_0,w_0,\mathbf{r},t)\,= \;\,\frac{\sigma_0}{2\pi}\,|k|\, \frac{(v\!-\!v_0)\,K_1(|k|\sqrt{(v\!-\!v_0)^2\!+\!(w\!-\!w_0)^2}\,)} {\sqrt{(v\!-\!v_0)^2\!+\!(w\!-\!w_0)^2}}\,e^{i(\omega t-ku)}
\label{Eq_v_comp_Rect_Magn_Field}\\
&\mathbf{e}_w\cdot\mathbf{h}_{\text{1D}}(\sigma_0,v_0,w_0,\mathbf{r},t) = \;\;\frac{\sigma_0}{2\pi}\,|k|\, \frac{(w\!-\!w_0)\,K_1(|k|\sqrt{(v\!-\!v_0)^2\!+\!(w\!-\!w_0)^2}\,)} {\sqrt{(v\!-\!v_0)^2\!+\!(w\!-\!w_0)^2}}\,e^{i(\omega t-ku)},
\label{Eq_w_comp_Rect_Magn_Field}
\end{align}
\end{subequations}
\end{widetext}
where $K_n$ denotes the $n$-th order modified Bessel function of the second kind.

With these expressions, we are equipped to calculate all elements of the dynamic demagnetizing tensor in the case where the magnetic cells are rectangular parallelepipeds. However, only a very few is actually needed to fully determine $\mathbf{n}^{(\alpha\beta)}$ under the assumption that the cells are arranged in a one-dimensional array, with the same $v$ coordinate. For such a cell array, indeed, even if all components of the dynamic magnetization effectively produce magnetic charges, four of the six off-diagonal elements systematically vanish. These are the $uv$, $vu$, $vw$, and $wv$ components. Taking also into account the requirements on the symmetry and trace of $\mathbf{n}^{(\alpha\beta)}$ mentioned in the introduction of Sec.~\ref{Sec_dynamic_Demag_Tensor}, one sees that no more than three dynamic demagnetization factors need to be known (only two if $\alpha\!=\!\beta$). Below, we derive mathematical expressions for a particular set of three factors, the diagonal elements $n_{uu}^{(\alpha\beta)}$ and $n_{ww}^{(\alpha\beta)}$ and the off-diagonal element $n_{wu}^{(\alpha\beta)}$, with which we can write $\mathbf{n}^{(\alpha\beta)}$ as
\begin{equation}
\mathbf{n}^{(\alpha\beta)}=
\begin{vmatrix}
\,&n_{uu}^{(\alpha\beta)}& &0& &n_{wu}^{(\alpha\beta)}&\\
\,&0& &\delta_{\alpha\beta}- \left(n_{uu}^{(\alpha\beta)}+n_{ww}^{(\alpha\beta)}\right)& &0&\\
\,&n_{wu}^{(\alpha\beta)}&& 0& &n_{ww}^{(\alpha\beta)}&\\
\end{vmatrix}.
\label{Eq_Rect_n}
\end{equation}

Let $w_{\beta}$ and $w_{\alpha}$ be the horizontal coordinates of the source and target cells, respectively. To calculate $n_{uu}^{(\alpha\beta)}$, we have to consider once again the $u$-component of the magnetic field produced by the volume charges in the source cell, that is here, all charge distributions $ikm_{0u}^{(\beta)}\,e^{i(\omega t-ku)}\,\delta(v\!-\!v_0)\,\delta(w\!-\!w_0)$ such that $-b/2\!\leqslant\!v_0\!\leqslant\!+b/2$ and $w_{\beta}\!-\!c/2\!\leqslant\!w_0\!\leqslant\!w_{\beta}\!+\!c/2$, and average it over the cross section of the target cell. This translates into
\begin{widetext}
\begin{equation}
-n_{uu}^{(\alpha\beta)}m_{u}^{(\beta)} = \frac{1}{bc}
\int_{w_{\alpha}-\frac{c}{2}}^{w_{\alpha}+\frac{c}{2}}\!
\int_{-\frac{b}{2}}^{+\frac{b}{2}}\!
\int_{w_{\beta}-\frac{c}{2}}^{w_{\beta}+\frac{c}{2}}\!
\int_{-\frac{b}{2}}^{+\frac{b}{2}}\!
\mathbf{e}_u\cdot\mathbf{h}_{\text{1D}}\!\left(\!ikm_{0u}^{(\beta)},v_0,w_0, \mathbf{r},t\right)\!dv_0\,dw_0\,dv\,dw,
\label{Eq_Rect_nuu1}
\end{equation}
and, after substituting Eq.~\ref{Eq_u_comp_Rect_Magn_Field} in Eq.~\ref{Eq_Rect_nuu1}, we get
\begin{equation}
n_{uu}^{(\alpha\beta)} = \frac{k^2}{2\pi bc}
\int_{w_{\alpha}-\frac{c}{2}}^{w_{\alpha}+\frac{c}{2}}\!
\int_{-\frac{b}{2}}^{+\frac{b}{2}}\!
\int_{w_{\beta}-\frac{c}{2}}^{w_{\beta}+\frac{c}{2}}\!
\int_{-\frac{b}{2}}^{+\frac{b}{2}}\!
K_0(|k|\sqrt{(v\!-\!v_0)^2\!+\!(w\!-\!w_0)^2}\,)
\,dv_0\,dw_0\,dv\,dw.
\label{Eq_Rect_nuu2}
\end{equation}
\end{widetext}
To calculate $n_{ww}^{(\alpha\beta)}$, we must consider this time the $w$-component of the magnetic field produced by the surface charges on the lateral faces of the source cell, i.e., all charge distributions $+m_{0w}^{(\beta)}\,e^{i(\omega t-ku)}\,\delta(v\!-\!v_0)\,\delta(w\!-\!c/2)$ (left face) and $-m_{0w}^{(\beta)}\,e^{i(\omega t-ku)}\,\delta(v\!-\!v_0)\,\delta(w\!+\!c/2)$ (right face) such that $v_{\beta}\!-\!b/2\!\leqslant\!v_0\!\leqslant\!v_{\beta}\!+\!b/2$, and average it over the target cell. This writes
\begin{widetext}
\begin{align}
-n_{ww}^{(\alpha\beta)}m_{w}^{(\beta)} = \frac{1}{bc}
\int_{w_{\alpha}-\frac{c}{2}}^{w_{\alpha}+\frac{c}{2}}\!
\int_{-\frac{b}{2}}^{+\frac{b}{2}}\!
\int_{-\frac{b}{2}}^{+\frac{b}{2}}
\mathbf{e}_w\,\cdot&\left[\,\mathbf{h}_{\text{1D}}\!\left(\!+m_{0w}^{(\beta)},v_0,+\frac{c}{2}, \mathbf{r},t\right)
+\mathbf{h}_{\text{1D}}\!\left(\!-m_{0w}^{(\beta)},v_0,-\frac{c}{2}, \mathbf{r},t\right)\right]
\!dv_0\,dv\,dw,
\label{Eq_Rect_nww1}
\end{align}
and, after substitution of Eq.~\ref{Eq_w_comp_Rect_Magn_Field} in Eq.~\ref{Eq_Rect_nww1}, we obtain
\begin{align}
n_{ww}^{(\alpha\beta)} = -\frac{|k|}{2\pi bc}
\int_{w_{\alpha}-\frac{c}{2}}^{w_{\alpha}+\frac{c}{2}}\!
\int_{-\frac{b}{2}}^{+\frac{b}{2}}\!
\int_{-\frac{b}{2}}^{+\frac{b}{2}}
&\left[\,\frac{(w\!-\!\frac{c}{2})\,K_1(|k|\sqrt{(v\!-\!v_0)^2\!+ \!(w\!-\!\frac{c}{2})^2}\,)} {\sqrt{(v\!-\!v_0)^2\!+\!(w\!-\!\frac{c}{2})^2}}
\right. \nonumber\\ &\left.
-\frac{(w\!+\!\frac{c}{2})\,K_1(|k|\sqrt{(v\!-\!v_0)^2\!+\!(w\!+\!\frac{c}{2})^2}\,)} {\sqrt{(v\!-\!v_0)^2\!+\!(w\!+\!\frac{c}{2})^2}}
\right]
\!dv_0\,dv\,dw.
\label{Eq_Rect_nww2}
\end{align}
Finally, calculating $n_{wu}^{(\alpha\beta)}$ requires to consider the $w$-component of the magnetic field produced by the volume charges $ikm_u^{(\beta)}$ in the source cell. Then, we have
\begin{equation}
-n_{wu}^{(\alpha\beta)}m_{u}^{(\beta)} = \frac{1}{bc}
\int_{w_{\alpha}-\frac{c}{2}}^{w_{\alpha}+\frac{c}{2}}\!
\int_{-\frac{b}{2}}^{+\frac{b}{2}}\!
\int_{w_{\beta}-\frac{c}{2}}^{w_{\beta}+\frac{c}{2}}\!
\int_{-\frac{b}{2}}^{+\frac{b}{2}}\!
\mathbf{e}_w\cdot\mathbf{h}_{\text{1D}}\!\left(\!ikm_{0u}^{(\beta)},v_0,w_0, \mathbf{r},t\right)\!dv_0\,dw_0\,dv\,dw,
\label{Eq_Rect_nwu1}
\end{equation}
and, after substituting Eq.~\ref{Eq_w_comp_Rect_Magn_Field} in Eq.~\ref{Eq_Rect_nwu1}, we find
\begin{equation}
n_{wu}^{(\alpha\beta)} = -i\;\frac{k|k|}{2\pi bc}
\int_{w_{\alpha}-\frac{c}{2}}^{w_{\alpha}+\frac{c}{2}}\!
\int_{-\frac{b}{2}}^{+\frac{b}{2}}\!
\int_{w_{\beta}-\frac{c}{2}}^{w_{\beta}+\frac{c}{2}}\!
\int_{-\frac{b}{2}}^{+\frac{b}{2}}
\frac{(w\!-\!w_0)\,K_1(|k|\sqrt{(v\!-\!v_0)^2\!+\!(w\!-\!w_0)^2}\,)} {\sqrt{(v\!-\!v_0)^2\!+\!(w\!-\!w_0)^2}}
\,dv_0\,dw_0\,dv\,dw.
\label{Eq_Rect_nwu2}
\end{equation}
\end{widetext}

We note that, using the same kind of reasoning, it would be rather straightforward to derive mathematical expressions for the redundant elements $n_{uw}^{(\alpha\beta)}$ and $n_{ww}^{(\alpha\beta)}$, and for all elements in the more general case of parallelepipedic cells arranged in a two-dimensional array (not considered here). We note also that the integrands and the integration domains are such that none of the multiple integrals that appear in the expressions of the dynamic demagnetization factors [Eqs.~\ref{Eq_Rect_nuu2},~\ref{Eq_Rect_nww2}, and~\ref{Eq_Rect_nwu2}] can be calculated analytically. Then, numerical methods must be employed to compute these factors. In cases where the source and target cells are totally disjoint, any interpolatory cubature rule (a multidimensional Simpson's rule in our case) may be used effectively. When they either coincide or when they share a face or an edge, however, one has to cope with the difficulty that the integrands have singularities at some points of the integration domains. In situations like these, Monte Carlo based algorithms such as the Vegas algorithm\,\cite{L78}, which is implemented in the GNU Scientific Library, are better suited. As a concluding remark to both Secs.~\ref{Sec_Slabs} and \ref{Sec_Parallelepipeds}, we note finally that the diagonal elements of the dynamic demagnetizating tensors are always real numbers whereas, as long as the magnetic cells are arranged in a one-dimensional array, the non-zero off-diagonal elements are systematically imaginary. A partial graphical explanation of why this is so will be given in Sec.~\ref{Sec_Discussion}.

\subsection{Discussion\label{Sec_Discussion}}
\subsubsection{Wave vector dependence of the dynamic demagnetization factors\label{Sec_k_Dependence}}
\begin{figure}[t]
\includegraphics[width=8.7cm,trim=25 163 25 130,clip]{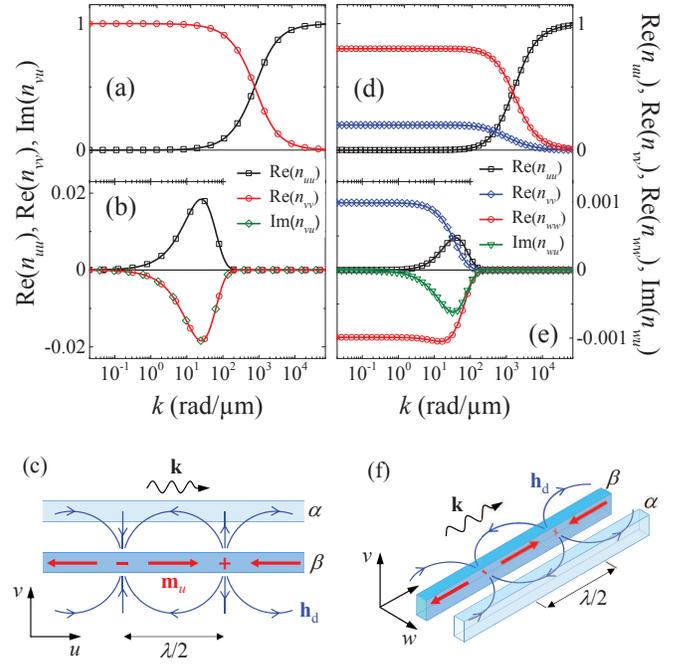}
\caption{(a),(b) $uu$, $vv$, and $vu$ components of the self (a) and mutual (b) dynamic demagnetizing tensors as a function of the wave vector $k$ for extended magnetic slabs of thickness $b\!=\!2$~nm separated by 40~nm ($v_{\alpha}\!>\!v_{\beta}$). (c),(f) Schematic representations of the dipole field, $\mathbf{h}_{\text{d}}$, created outside a source cell ($\beta$) by the $u$-component of the variable magnetization in a plane spin-wave propagating along $u$, for slabs (c) and parallepipeds (f). (d),(e) $uu$, $ww$ and $wu$ components of the self (d) and mutual (e) dynamic demagnetizing tensors as a function of the wave vector $k$ for magnetic parallepipeds of height $b\!=\!5$~nm and width $c\!=\!2$~nm, separated by 40~nm ($w_{\alpha}\!>\!w_{\beta}$).} \label{Fig_Dynamic_Demag_Factors}
\end{figure}

As illustrated in Fig.~\ref{Fig_Dynamic_Demag_Factors}(a,b), the $uu$ elements of the self and mutual dynamic demagnetizing tensors of extended magnetic slabs show very different variations with $k$. The self demagnetization factor varies monotonously [Fig.~\ref{Fig_Dynamic_Demag_Factors}(a)]. It goes from zero at low wave vector, where the magnetic charges $-\boldsymbol\nabla\!\cdot\!
\mathbf{m}^{(\beta)}\!=\!i\,k\,m_{u}^{(\beta)}$ created by $\mathbf{m}_u^{(\beta)}\!=\!m_u^{(\beta)}\mathbf{e}_u$ are extremely dilute, to unity at high wave vectors, where successive spin-wave wavefronts are so close to one another that these volumes charges become spatially distributed very much like surface charges in an ensemble of closely packed, perpendicularly magnetized thin magnetic films, sitting vertically. The transition between the two regimes occurs for $k\!\simeq\!b^{-1}$. In contrast, the mutual demagnetization factor vanishes both in the low wave vector limit (for the same reason as the self demagnetization factor) and in the high wave vector limit [Fig.~\ref{Fig_Dynamic_Demag_Factors}(b)]. In the latter case, this occurs because the alternated positive and negative charges associated with the spatial variation of $\mathbf{m}_u^{(\beta)}$ become indistinguishable, as viewed from the target cell, and average out to zero. In between these two limits [Fig.~\ref{Fig_Dynamic_Demag_Factors}(c)], the magnetic charges are arranged so that the dynamic dipolar coupling between the source and target cells is sizeable. $n_{uu}^{(\alpha\beta)}$ reaches a maximum value for $k\!=\!|v_{\alpha}-v_{\beta}|^{-1}$, which is inversely proportional to the cell separation and approximately amounts to $\frac{1}{2}b |v_{\alpha}-v_{\beta}|^{-1}$. Figure~\ref{Fig_Dynamic_Demag_Factors}(c) illustrates the reason why the $vu$ element of the mutual demagnetizing tensor is imaginary while the $uu$ one is real [Eqs.~\ref{Eq_Slab_nuu_Mutual}, \ref{Eq_Slab_nuv_Mutual_bis}]: this reflects the fact that the $u$ and $v$ components of the dipole field created by $\mathbf{m}_u^{(\beta)}$ at the location of the target cell oscillate in quadrature and are maximum in places separated by a quarter of spin-wave wavelength.

Moving to parallepipeds [Fig.~\ref{Fig_Dynamic_Demag_Factors}(f)], that is,  limiting the size of the magnetic cells in both dimensions perpendicular to the direction of spin-wave propagation, induces additional finite size effects as compared to extended slabs. For the description of the changes produced, which are qualitative as well as quantitative, one should keep in mind that moving from our film geometry [Fig.~\ref{Fig_Magnetic_Cells}(a)] to our strip geometry [Fig.~\ref{Fig_Magnetic_Cells}(b)] also requires to rotate the magnetic medium by $90^{\text{o}}$ about axis $u$, i.e., to interchange $v$ and $w$. First of all, the factors related to the newly confined dimension (the self and mutual $n_{vv}$) are no longer systematically nil [Fig.~\ref{Fig_Dynamic_Demag_Factors}(d,e)]. This obviously follows from the creation of some dipole field by the so far "silent" component of the variable magnetization ($m_w$ in the film geometry). Second, the $k$-dependence of some of the factors is altered, especially in the low-$k$ limit. For instance, the self demagnetization factor $n_{ww}$ (the pendant of $n_{vv}$ in the film geometry) tends towards a value, which is reduced below unity [Fig.~\ref{Fig_Dynamic_Demag_Factors}(d)] all the more strongly, the smaller the aspect ratio $b/c$. Concomitantly, its mutual counterpart, while still exhibiting a local optimum, takes on a non-zero value when $k\rightarrow 0$ [Fig.~\ref{Fig_Dynamic_Demag_Factors}(e)]. Third, as a comparison of the vertical scales in panels (b) and (e) of Fig.~\ref{Fig_Dynamic_Demag_Factors} reveals, the reduction of the second lateral cell dimension, at fixed cell separation, is accompanied by a global decrease of the magnitude of all the mutual dynamic demagnetization factors. This is the natural consequence of an increased spatial dilution of the dynamic stray field produced by the source cell as the target cell subtends a smaller solid angle [Fig.~\ref{Fig_Dynamic_Demag_Factors}(f)].

\subsubsection{Validation of the method\label{Sec_Validity}}
\begin{figure}[t]
\includegraphics[width=8.5cm,trim=0 60 0 50,clip]{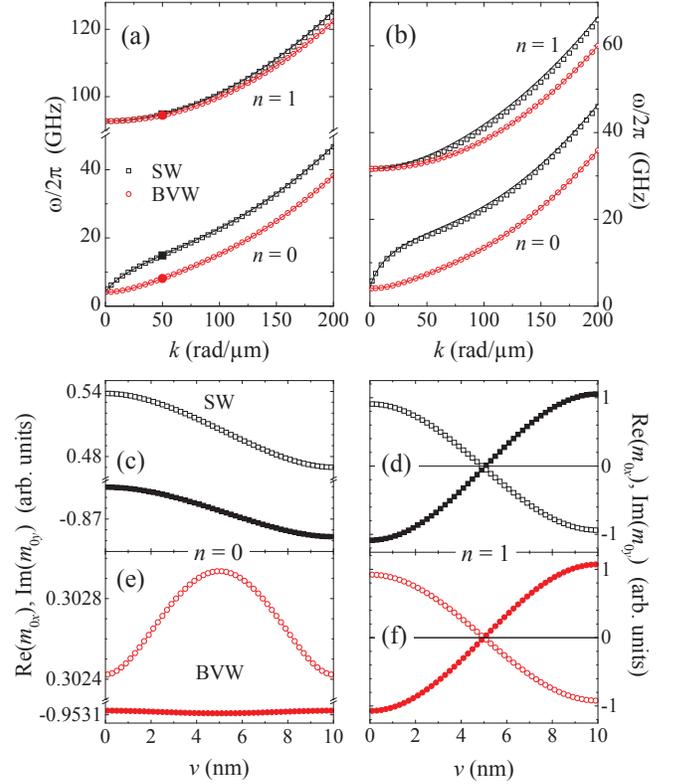}
\caption{(a),(b) Dispersion relations of the first two spin-wave branches in 10~nm (a) and 20~nm (b) thick films with $A\!=\!11$~pJ/m and $M_{\text{S}}\!=\!800$~kA/m. Symbols and lines correspond to results of our numerical approach ($b=0.2$~nm) and predictions of the zero order perturbation theory of Kalinikos and Slavin [Eq.~\ref{Eq_KS}], respectively. The films are magnetized in-plane, along $u$ (red circles) or $w$ (black squares). $\mu_0H_0\!=\!20$~mT. (c)-(f) $v$-profiles of the $n\!=\!0$ (c,e) and $n\!=\!1$ (d,f) backward volume wave modes (c,d) and surface wave modes (e,f) with $k\!=\!50$~rad/$\mu$m in the 10~nm thick film [indicated with solid symbols in (a)]. Open and solid symbols represent the out-of-plane ($y\!=\!v$) and in-plane components of the dynamic magnetization, respectively. In (c,d) $x=-w$. In (e,f) $x=u$.}\label{Fig_KS}
\end{figure}

As reported in Appendix~\ref{App_Validation}, a number of tests have been performed in order to check the validity of our theoretical results regarding the plane-wave demagnetization factors [Secs.~\ref{Sec_Slabs} and \ref{Sec_Parallelepipeds}]. None of these tests consisting in comparisons with results from analytical models and examination of various limiting cases has proved our results wrong. As an introduction to the use of our method for exploring propagating spin-wave physics, we describe hereafter another demanding test, which demonstrates the validity of our numerical scheme. In this test, dispersion relations computed for anisotropy-free homogeneous extended films in the dipole-exchange regime are compared to predictions of the perturbation theory of Kalinikos and Slavin\,\cite{KS86}. The latter uses the magnetostatic Green's function method to account for dipolar interactions and allows for the derivation of explicit, though approximate, expressions for the dispersion relations\,\cite{K81}. At order zero in perturbation and in the absence of surface pinning, the dispersion relations of the $n$-th mode in the backward volume wave (BVW, $\mathbf{M}_{\text{eq}}\!\parallel\!u$) and surface wave (SW, $\mathbf{M}_{\text{eq}}\!\parallel\!w$) configurations are
\begin{subequations}\label{Eq_KS}
\begin{align}
(\text{BVW})\;\;\,
\omega_n^2 =& \left(\omega_{\text{H}}+\omega_{\text{M}}\Lambda^2k_n^2\right)
\nonumber\\ & \times
\left(\omega_{\text{H}}+\omega_{\text{M}}\Lambda^2k_n^2+\omega_{\text{M}}(1-P_{nn})\right)
\label{Eq_KS_BVW}\\
(\text{SW})\;\;\,
\omega_n^2 =& \left(\omega_{\text{H}}+\omega_{\text{M}}\Lambda^2k_n^2+\omega_{\text{M}}P_{nn}\right)
\nonumber\\ & \times
\left(\omega_{\text{H}}+\omega_{\text{M}}\Lambda^2k_n^2+\omega_{\text{M}}(1-P_{nn})\right)
\label{Eq_KS_SW}
\end{align}
\end{subequations}
where $n$ is the quantization number along the film thickness, $\Lambda$ still denotes the exchange length, $k_n^2=k^2+(n\pi/T)^2$, and $P_{nn}$ is given by
\begin{equation}
P_{nn} = \frac{k^2}{k_n^2}
\left[
1 - \left(\frac{2}{1+\delta_{0n}}\right) \frac{k^2}{k_n^2} \left(\frac{1-(-1)^ne^{-|k|T}}{|k|T}\right)
\right].
\label{Eq_Pnn}
\end{equation}
Figure~\ref{Fig_KS}(a) illustrates the fact that, as long as the film thickness remains moderate, the dispersion relations of the first two modes ($n\!=\!0,1$) computed with our finite-difference approach match those calculated with Eqs.~\ref{Eq_KS} quite closely, for both magnetic configurations. This is a proof that our numerical scheme is correct, not only as far as dipolar interactions are concerned, but also regarding how the exchange interactions are treated.

Some clear discrepancy however appears at large film thickness [Fig.~\ref{Fig_KS}(b)]. This does not come as a surprise since, on increasing $T$, the frequency distance between the spin-wave branches with $n\!=\!0$ and $n\!=\!1$ decreases so that their dipole-dipole hybridization may become significant, which is not accounted for by the zero-order approximation\,\cite{KS86}. Noticeably, deviations from the computed data are observed earlier for surface waves\,\cite{Rem5} than for volume waves. This reduction of the thickness range of applicability of the analytical model for surface waves is likely related to their specific modal profile. Unlike volume waves, which have well defined profile symmetry [Fig.~\ref{Fig_KS}(e,f)], surface waves are neither fully symmetric nor fully antisymmetric [Fig.~\ref{Fig_KS}(c,d)]. As a consequence, hybridization between branches of odd and even indices, which is not permitted for volume waves, is allowed for surface waves and hybridization is thus generally stronger in the SW configuration. We have checked that on including explicitly the hybridization between the $n\!=\!0$ and $n\!=\!1$ surface wave branches, as is done for example in Refs.~\onlinecite{HBKL14} and \onlinecite{GHHK16}, the data produced by the analytical model are lying significantly closer to those computed with our numerical approach (not shown).

\section{Applications\label{Sec_Applications}}
The physical situations where the finite-difference approach that we propose should prove most useful are either those where the material parameters vary throughout the magnetic medium or those where the medium is not homogeneously magnetized, two possibilities that are difficult to include in an analytical spin-wave theory such as the one developed for films\,\cite{KS86,K81}. In order to illustrate this point and, simultaneously, give examples of application of our numerical model in the two geometries considered here, we will address below two questions of current interest: i) the frequency non-reciprocity of surface waves in films with heterogenous magnetic properties (Sec.~\ref{Sec_Films}) and ii) the channeling of spin-waves inside magnetic domain walls (Sec.~\ref{Sec_Strips}). It should be noticed that all the non-collinear equilibrium spin configurations discussed hereafter have been determined by solving numerically overdamped Landau-Lifshitz-Gilbert equations with a fourth order Runge-Kutta method.

\begin{figure}[t]
\includegraphics[width=8.5cm,trim=0 130 0 110,clip]{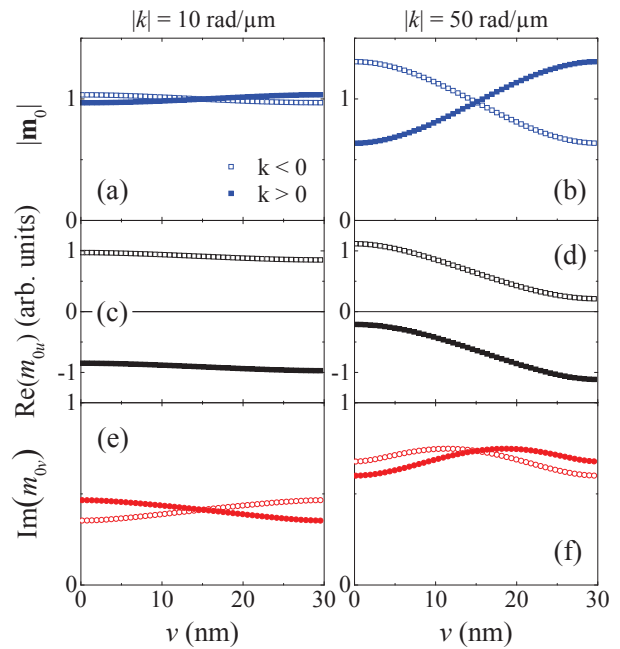}
\caption{Profiles of the fundamental surface wave modes with $k\!>\!0$ (solid symbols) and $k\!<\!0$ (open symbols) in a 30~nm thick film of permalloy ($A\!=\!11$~pJ/m, $M_{\text{S}}\!=\!800$~kA/m), for $|k|\!=\!10$~rad/$\mu$m (left column) and $|k|\!=\!50$~rad/$\mu$m (right column): (a),(b) total amplitude, (c),(d) in-plane component and (e),(f) out-of-plane component of the dynamic magnetization. The external magnetic field is $\mu_0H_0\!=\!50$~mT. All plotted quantities are in arbitrary units.}
\label{Fig_Localization}
\end{figure}

\begin{figure}[t]
\includegraphics[width=8.2cm,trim=0 95 0 80,clip]{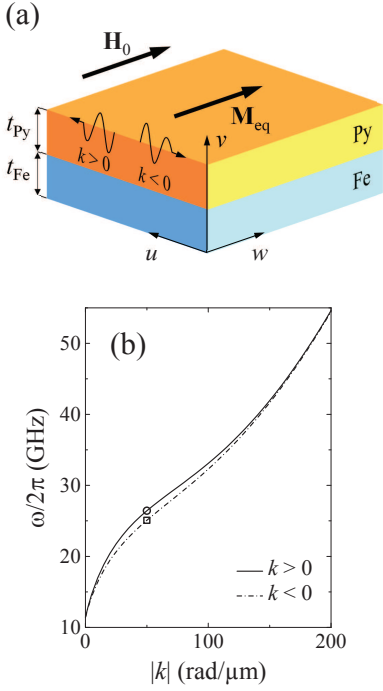}
\caption{(a) Schematic representation of a transversally magnetized (001)Fe/Py bilayer film. (b) Dispersion relation of the fundamental surface wave mode in a film with $t_{\text{Fe}}\!=\!t_{\text{Py}}\!=\!7.5$~nm submitted to an external magnetic field $\mu_0H_0\!=\!50$~mT, for
positive (solid line) and negative (dashed line) wave vectors. The symbols indicate the modes whose $v$-profiles are shown in Fig.~\ref{Fig_Bilayer_Modes}.}
\label{Fig_Bilayer_Dispersion}
\end{figure}

\subsection{Inhomogeneous magnetic films\label{Sec_Films}}
Owing to their largest group velocity, surface waves ($\mathbf{M}_{\text{eq}}\!\parallel\!w$) are often considered as the most relevant spin-waves for magnonic applications\,\cite{K13}. They are also special in that they are the only standard spin-waves for which i) two components of the dynamic dipole field $\mathbf{h}_{\text{d}}$, one in-plane ($u$) and one out-of-plane ($v$), contribute to the torque acting on the dynamic magnetization and ii) the off-diagonal elements of the mutual demagnetizing tensor [Eq.~\ref{Eq_Slab_n_Mutual}], which change sign on reversing the direction of propagation, play an important role. As illustrated in Fig.~\ref{Fig_Localization}(a,b), these peculiarities lead to the formation of asymmetric distributions of dynamic magnetization across the film thickness, such that waves propagating in opposite directions have larger amplitudes near opposite surfaces. Because of this specific character, also, counter-propagating spin-waves of a given wave vector $|k|$ have different frequencies as soon as the film exhibits vertically asymmetric properties like, for instance, inequivalent magnetization pinning (anisotropy) at the top and bottom surfaces\,\cite{H90,GHHK16}. Our numerical approach is particularly well suited to compute the frequency non-reciprocities produced by all sorts of magnetic symmetry breaking. Here, we will consider the case of a bilayer film made of two ferromagnetic materials with different exchange stiffness and saturation magnetization.

\begin{figure}[t]
\includegraphics[width=8.5cm,trim=0 130 0 110,clip]{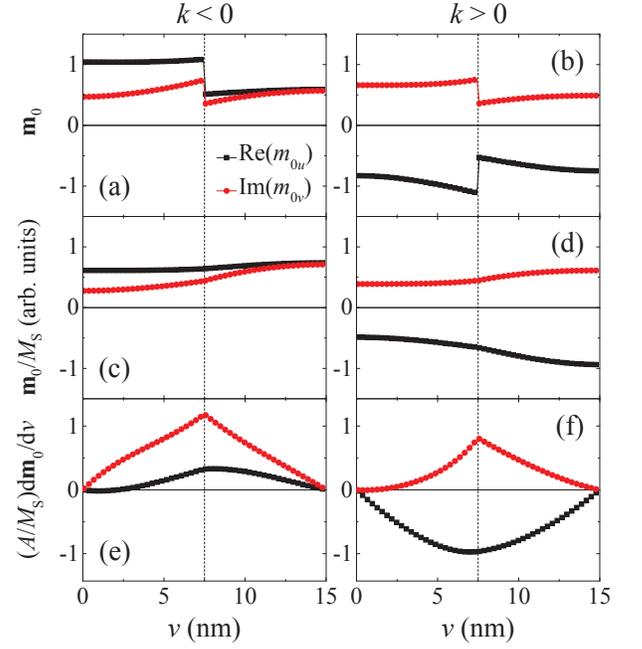}
\caption{(a),(b) $v$-profiles of the fundamental surface wave modes with $|k|\!=\!50$~rad/$\mu$m through a Fe/Py bilayer film with $t_{\text{Fe}}\!=\!t_{\text{Py}}\!=\!7.5$~nm submitted to an external field $\mu_0H_0\!=\!50$~mT. (c)-(f) Variations of the quantities $\frac{\mathbf{m}_0}{M_{\text{S}}}$ (c,d) and  $\frac{A}{M_{\text{S}}}\frac{\partial\mathbf{m}_0}{\partial v}$ (e,f) with the $v$-coordinate, as deduced from the mode profiles shown in (a) and (b). In each panel, the in-plane ($u$) and out-of-plane ($v$) components are shown as black squares and red circles, respectively. The left and right columns correspond to modes with $k\!<\!0$ ($f\!=\!20.0$~GHz) and $k\!>\!0$ ($f\!=\!21.1$~GHz), respectively. All plotted quantities are in arbitrary units.}
\label{Fig_Bilayer_Modes}
\end{figure}

Before proceeding with the description of our results, a technical remark must be made. When $A$ and $M_{\text{S}}$ vary in space, the exchange interaction must be treated carefully. Its contributions to the equilibrium and dynamic magnetic fields [Eqs.~\ref{Eq_Hex2} and \ref{Eq_hex1}] can no longer be expressed in terms of exchange length, which is a concept only valid inside a homogeneous magnetic material. New expressions must be used, where $A$ and $M_{\text{S}}$ appear explicitly. Starting from the Heisenberg formulation of the exchange energy and assuming that the angle between adjacent spins remain small, one may easily show that Eq.~\ref{Eq_Hex2} becomes
\begin{align}
\mathbf{H}_{\text{ex}}^{(\alpha)} =& \frac{2A^{(\alpha-)}}{\mu_0M_{\text{S}}^{(\alpha)}\xi^2}
\left(\frac{\mathbf{M}_{\text{eq}}^{(\alpha-1)}}{M_{\text{S}}^{(\alpha-1)}} -
\frac{\mathbf{M}_{\text{eq}}^{(\alpha)}}{M_{\text{S}}^{(\alpha)}}\right) (1\!-\!\delta_{1\alpha})
\nonumber\\
&+\frac{2A^{(\alpha+)}}{\mu_0M_{\text{S}}^{(\alpha)}\xi^2}
\left(\frac{\mathbf{M}_{\text{eq}}^{(\alpha+1)}}{M_{\text{S}}^{(\alpha+1)}} -
\frac{\mathbf{M}_{\text{eq}}^{(\alpha)}}{M_{\text{S}}^{(\alpha)}}\right) (1\!-\!\delta_{N\alpha}),
\label{Eq_Hex3}
\end{align}
where $A^{(\alpha\pm)}$ denotes the value of the exchange coefficient between cell $\alpha$ and cell $\alpha\!\pm\!1$, which we choose here to express as the harmonic mean of the exchange stiffness constants in the volume of the cells, $A^{(\alpha\pm)}\!=\!\frac{2A^{(\alpha)}A^{(\alpha\pm1)}} {A^{(\alpha)}+A^{(\alpha\pm1)}}$. Similarly, Eq.~\ref{Eq_hex1} becomes
\begin{align}
\mathbf{h}_{\text{ex}}^{(\alpha)} = &\frac{2A^{(\alpha-)}}{\mu_0M_{\text{S}}^{(\alpha)}\xi^2}
\left(\frac{\mathbf{m}^{(\alpha-1)}}{M_{\text{S}}^{(\alpha-1)}} -
\frac{\mathbf{m}^{(\alpha)}}{M_{\text{S}}^{(\alpha)}}\right) (1\!-\!\delta_{1\alpha})
\nonumber\\
&+\frac{2A^{(\alpha+)}}{\mu_0M_{\text{S}}^{(\alpha)}\xi^2}
\left(\frac{\mathbf{m}^{(\alpha+1)}}{M_{\text{S}}^{(\alpha+1)}} -
\frac{\mathbf{m}^{(\alpha)}}{M_{\text{S}}^{(\alpha)}}\right) (1\!-\!\delta_{N\alpha})
\nonumber\\
&-\frac{2A^{(\alpha)}}{\mu_0\left.M_{\text{S}}^{(\alpha)}\right.^2}\;k^2 \;\mathbf{m}^{(\alpha)},
\label{Eq_hex3}
\end{align}
and equations~\ref{Eq_hex2} should be modified accordingly.

\begin{figure*}[t]
\includegraphics[width=17cm,trim=20 160 10 175,clip]{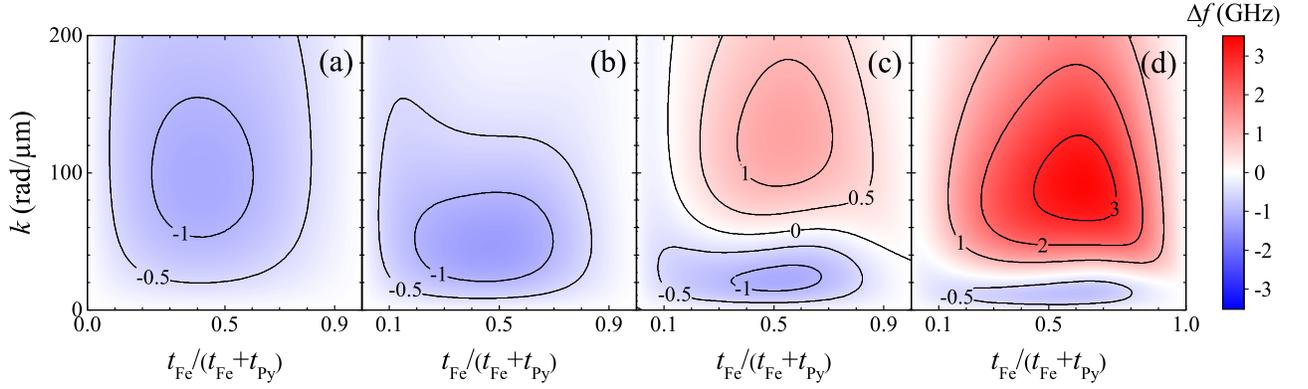}
\caption{Frequency non-reciprocity $\Delta f\!=\!f(-|k|)\!-\!f(|k|)$ of the fundamental surface wave mode as a function of the wave vector $k$ and composition, for Fe/Py bilayer films of varying total thickness $T$: (a) $T\!=\!10$~nm, (b) $T\!=\!15$~nm, (c) $T\!=\!20$~nm, and (d) $T\!=\!25$~nm  ($\mu_0H_0\!=\!50$~mT).}
\label{Fig_Frequency_NR_Maps}
\end{figure*}

\begin{figure}[b]
\includegraphics[width=8.5cm,trim=60 115 120 115,clip]{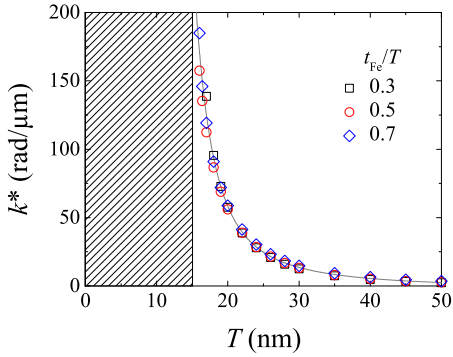}
\caption{Wave vector $k^{\ast}$ at which the frequency non-reciprocity $\Delta f$ of the fundamental surface wave mode in Fe/Py bilayer films changes sign as a function of the film thickness $T$, for three values of the relative fraction of Fe, $t_{\text{Fe}}/T\!=\!0.3$ (squares), $t_{\text{Fe}}/T\!=\!0.5$ (circles), and $t_{\text{Fe}}/T\!=\!0.7$ (diamonds). The shaded zone indicates the thickness range where no change of sign occurs. The line is a guide to the eye. $\mu_0H_0=50$~mT. }
\label{Fig_kstar_vs_T}
\end{figure}

\begin{figure}[b]
\includegraphics[width=8.6cm,trim=55 70 55 75,clip]{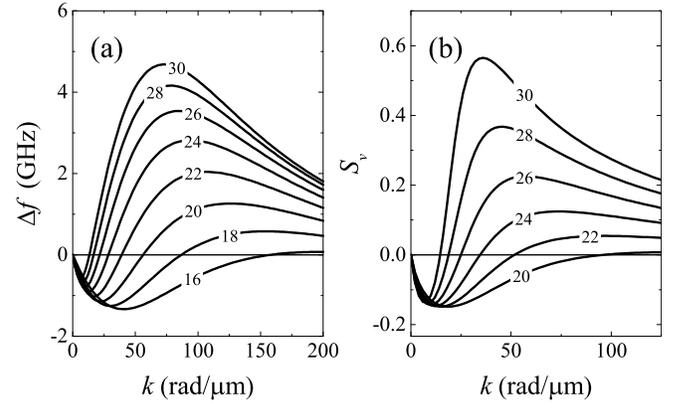}
\caption{(a) Variation of the frequency non-reciprocity $\Delta f = f(-|k|)-f(|k|)$ of the fundamental surface wave mode as a function of the wave vector $k$ in Fe/Py bilayer films with thickness $T$ varying from 16~nm (bottommost curve) to 30~nm (topmost curve). All bilayers are such that $t_{\text{Fe}}/T\!=\!0.5$. (b) Variation of the asymmetry parameter $S_v$ with $k$ in homogenous films of thickness $T$ varying from 20~nm (bottommost curve) to 30~nm (topmost curve). The films are supposed to be made of a hypothetical material with mean magnetic parameters $A=15.5$~pJ/m and $M_{\text{S}}\!=\!1250$~kA/m. The data shown in panels (a) and (b) do not correspond one-to-one. They have been chosen so as to best reveal the similitude between the behaviors of the two quantities, $\Delta f$ and $S_v$.}
\label{Fig_Sv}
\end{figure}

The system we consider now consists of a permalloy (Py) layer ($A\!=\!11$~pJ/m, $M_{\text{S}}\!=\!800$~kA/m) of thickness $t_{\text{Py}}$ lying on top of and exchange coupled to a single crystal \textit{bcc} Fe layer ($A\!=\!20$~pJ/m, $M_{\text{S}}\!=\!1700$~kA/m, $K_\text{c}\!=\!50$~kJ/m$^3$) of thickness $t_{\text{Fe}}$ [Fig.~\ref{Fig_Bilayer_Dispersion}(a)]. The Fe crystal is oriented so that $\{\mathbf{c}_1,\mathbf{c}_2, \mathbf{c}_3\} = \{\mathbf{e}_u,\mathbf{e}_v,\mathbf{e}_w\}$ and the external magnetic field is applied parallel to $\mathbf{e}_w\!=\!\mathbf{c}_3$, which is an easy direction of magnetization for the Fe component, in order to magnetize the film at right angle to the propagation direction, $\{x,y,z\}\!=\{u,v,w\}$. As may be seen in Fig.~\ref{Fig_Bilayer_Modes}(a,b), the bi-component character of the film strongly manifests itself in the profile of the normal modes, in the form of discontinuities at the location of the Fe/Py interface. As expected, the ratios $\text{Re}(m_{0u})/M_{\text{S}}$ and $\text{Im}(m_{0v})/M_{\text{S}}$, which are measures of the precession angles of the magnetization, remain continuous there but they exhibit clear changes of slope [Fig.~\ref{Fig_Bilayer_Modes}(c,d)]. The latter are necessary to fulfil the micromagnetic boundary condition\,\cite{MD07}, which requires that $\frac{A}{M_{\text{S}}}\frac{\partial\mathbf{m}_0} {\partial v}$ be continuous across the interface [Fig.~\ref{Fig_Bilayer_Modes}(e,f)].

As illustrated in Fig.~\ref{Fig_Bilayer_Dispersion}(b), a difference in the frequencies of counter-propagating spin-waves is observed as soon as the wave vector is not zero and the film is indeed magnetically asymmetric ($t_{\text{Py}}t_{\text{Fe}}\!\neq\!0$). Noticeably, the frequency non-reciprocity effect appears as maximum for $t_{\text{Py}}/t_{\text{Fe}}$ of order unity, irrespective of the total film thickness $T\!=\!t_{\text{Py}}+t_{\text{Fe}}$ [Fig.~\ref{Fig_Frequency_NR_Maps}]. Nevertheless, a rich behavior is observed when varying $T$. While in thin films ($T\!\leqslant\!15$~nm), the frequency difference $\Delta f = f(-|k|)-f(|k|)$ for the fundamental SW mode is always negative [Fig.~\ref{Fig_Frequency_NR_Maps}(a,b)], in thick films, it goes from negative to positive with increasing $k$ [Fig.~\ref{Fig_Frequency_NR_Maps}(c,d)]. This sign reversal occurs for a wave vector $k^{\ast}$ which decreases fast with increasing $T$ [Fig.~\ref{Fig_kstar_vs_T}] but is only weakly dependent on the film composition [Figs.~\ref{Fig_Frequency_NR_Maps}(d) and \ref{Fig_kstar_vs_T}], at least in the range $0.2\!<\!t_{\text{Fe}}/T\!<\!0.8$. This is an indication that the change of sign of $\Delta f$ is not due to the magnetic asymmetry itself. It is rather related to an intrinsic phenomenon, which occurs also in symmetric films and just gets highlighted when the magnetic symmetry is broken. We note that, in thick films with modal profiles not perturbed by any kind of magnetic asymmetry, the overall localization of the fundamental SW mode (looking at $|\mathbf{m}_0|$) does not reverse when $k$ increases [Fig.~\ref{Fig_Localization}(a,b)] but the side of the film where the out-of-plane component of the dynamic magnetization ($m_{0v}$) is the largest does [Fig.~\ref{Fig_Localization}(e,f)]. Furthermore, the parameter $S_v\!=\!\text{Im}\left(\frac{m_{0v}(0)\,-\,m_{0v}(T)} {m_{0v}(0)\,+\;m_{0v}(T)}\right)$, which measures the intrinsic degree of asymmetry of the profile of $m_{0v}$ in homogeneous films, varies with $k$ in the same qualitative manner as $\Delta f$ does for composite films [Fig.~\ref{Fig_Sv}]. Then it seems that the behavior of the frequency non-reciprocity in bilayer films is somehow related to that of $m_{0v}$. It is however beyond the scope of the present paper to elucidate why this is so. For that, a dedicated analytical theory would certainly be necessary, such as the one developed in Ref.~\onlinecite{GHHK16} to account for the effect of a difference in anisotropy at the two films surfaces.

\begin{figure}[t]
\includegraphics[width=8.3cm,trim=20 85 15 60,clip]{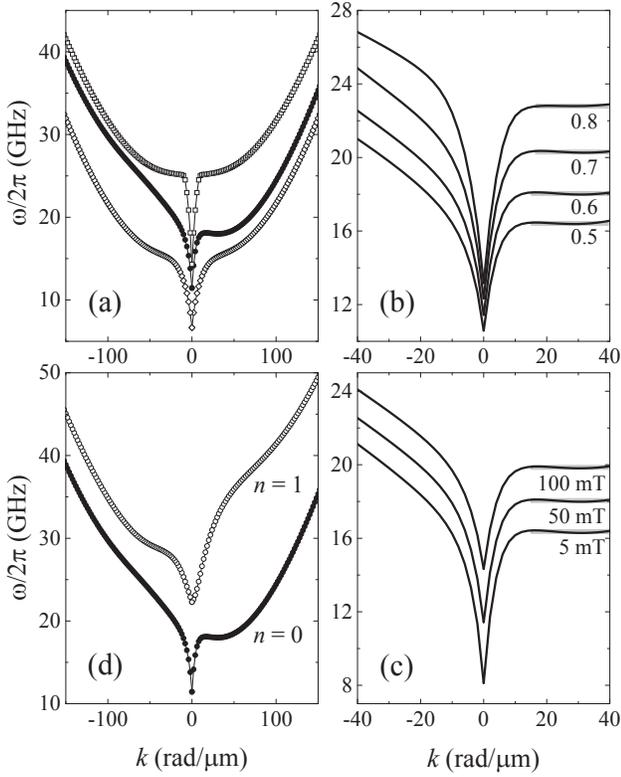}
\caption{Dispersion relations of surface wave modes in 34~nm thick films. (a) Fundamental mode in pure Fe (open squares) and Py (open diamonds) films, and in a Fe/Py bilayer with $t_{\text{Fe}}/T=0.6$ (solid circles), in a transverse magnetic field $\mu_0H_0\!=\!50$~mT. (b) Fundamental modes in Fe/Py bilayers with Fe contents $t_{\text{Fe}}/T$ varying from 0.5 to 0.8 ($\mu_0H_0\!=\!50$~mT). (c) Fundamental mode in a Fe/Py bilayer with $t_{\text{Fe}}/T\!=\!0.6$ submitted to different magnetic fields. (d) First two SW modes in a Fe/Py bilayer with $t_{\text{Fe}}/T=0.6$ ($\mu_0H_0\!=\!50$~mT).}
\label{Fig_SW_Filter}
\end{figure}

In thick films, the overall magnitude of the frequency non-reciprocity effect increases monotonously with increasing $T$ [Figs.~\ref{Fig_Frequency_NR_Maps} and \ref{Fig_Sv}(a)]. This originates essentially from the combination of two factors: i) the larger the film thickness the larger the intrinsic modal profile asymmetry and ii) the larger the modal profile asymmetry the larger $\Delta f$, for a given magnetic asymmetry. Thus, for $T\!\geqslant\!30$~nm, extremely large effects with $\Delta f$s of several GHz can be obtained. They are associated with dispersion relations, which fall in between those of pure Fe and Py films of the same thickness [Fig.~\ref{Fig_SW_Filter}(a)], but end up being extremely asymmetric because each of their positive and negative-$k$ branches tends to follow the $\omega(k)$ curve of the material where the mode is more strongly localized: the negative-$k$ branch is pulled upwards, towards the $\omega(k)$ curve of Fe, while the positive-$k$ branch is pulled downwards, towards that of Py. As a byproduct of this skewing, the dispersion relations of thick Fe/Py bilayers quite systematically show a well defined frequency plateau, that is, a range of positive $k$ values where the group velocity $v_{\text{g}}\!= \!\frac{\partial\omega} {\partial k}$ is close to zero. For spin-waves of the corresponding frequencies, effective propagation is only possible with a negative wave vector, i.e., in the $-u$ direction ($\omega\!>\!0$). The narrow frequency window in question can be widely tuned by changing the composition of the film [Fig.~\ref{Fig_SW_Filter}(b)] or the magnitude of the external magnetic field [Fig.~\ref{Fig_SW_Filter}(c)]. Such an usual behavior might be useful in applications, for instance, to build narrow band microwave isolators. Of course, all the non-reciprocity phenomena discussed above switch sign or invert when $\mathbf{H}_0$ is reversed and the bilayer film is magnetized along $-w$. As may be seen in Fig.~\ref{Fig_SW_Filter}(d), frequency non-reciprocities also qualitatively invert when moving from the first SW mode ($n\!=\!0$) to the second ($n\!=\!1$).

\subsection{Inhomogeneously magnetized strips\label{Sec_Strips}}
As demonstrated recently using time-domain micromagnetic simulations\,\cite{GBSA15} and even more recently through Brillouin light scattering experiments\,\cite{WKSH16}, a magnetic domain wall can act as a magnonic waveguide. The reasons for this are essentially twofold. First, a domain wall quite systematically hosts a spin-wave mode, which is strongly localized sidewise by the confining potential of the magnetic texture but free to propagate in the direction parallel to the wall. Second, this bound spin-wave mode lies in the energy gap of the usual extended (bulk) spin-wave modes and is therefore spectrally isolated, at least at low $k$. This remarkable ability of domain walls to channel spin-waves provides an efficient solution to the difficult problem of guiding spin-waves along curved and/or reprogramable paths\,\cite{GBSA15,WKSH16,LYWX15}. It is believed that it could play a crucial role in the future development of magnonic circuits.

\begin{figure}[t]
\includegraphics[width=8.65cm,trim=35 200 35 155,clip]{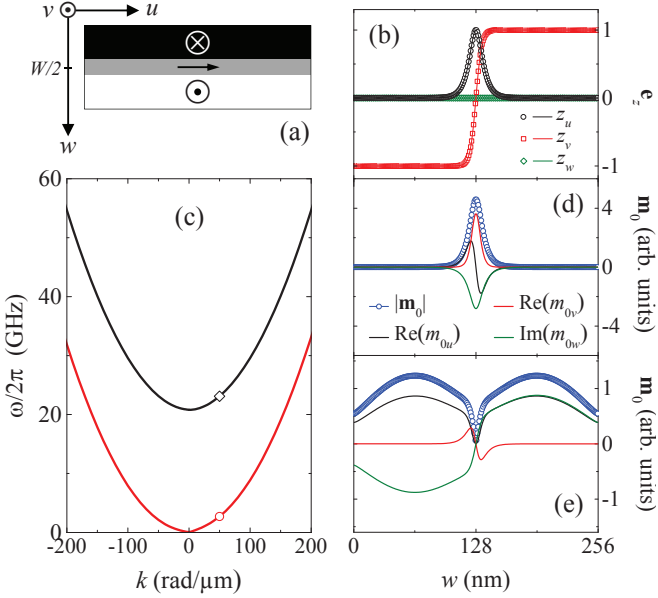}
\caption{(a) Schematic representation of a magnetic strip with perpendicular-to-plane anisotropy containing a single Bloch wall in its centre. (b) Variation of the equilibrium magnetization direction $\mathbf{e}_z$ ($z_i=\mathbf{e}_z\!\cdot\!\mathbf{e}_i$, $i=u,v,w$) across such a strip, 1~nm thick and 256~nm wide. Results of numerical simulations (symbols) are compared to predictions of Walker's analytical model with $p_u\!=\!p_v\!=\!+1$ and $\Delta_{\text{w}}\!=\!6.1$~nm (lines), see text for details. (c) Dispersion relations of the propagating spin-wave normal modes of lowest and second lowest frequencies in the magnetic configuration shown in (b). (d),(e) $w$-profiles of these two modes at the points marked with symbols in (c), i.e., for $k=+50$~rad/$\mu$m ($x=-w$).}
\label{Fig_DW_Profile}
\end{figure}

Below, we use our dynamic matrix approach to study domain-wall channelized spin-wave (DWCSW) normal modes. We consider the case of Bloch walls formed in an hypothetical material with strong perpendicular-to-plane uniaxial magnetic anisotropy ($A\!=\!15$~pJ/m, $M_{\text{S}}\!=\!1$~MA/m, $K_\text{u}\!=\!1$~MJ/m$^3, \;\mathbf{a}\!=\!\mathbf{e}_v$)\,\cite{GBSA15}. We examine the situation where a unique and straight wall runs along the entire length of a magnetic strip, dividing it laterally in two oppositely magnetized domains [Fig.~\ref{Fig_DW_Profile}(a,b)]. We assume zero external magnetic field so that the wall sits at the center of the strip. For $W\!\gg\!b$, the equilibrium magnetic configuration follows Walker's profile\,\cite{SW74}, i.e., $z_w\!=\!\mathbf{e}_z\!\cdot\!\mathbf{e}_w\!=\!0$, $z_v\!=\!\mathbf{e}_z\!\cdot\!\mathbf{e}_v\!= \!p_v\tanh\!\left(\frac{w-w_0}{\Delta_{\text{w}}}\right)$, $z_u\!=\!\mathbf{e}_z\!\cdot\!\mathbf{e}_u\!= \!p_u\sech\!\left(\frac{w-w_0}{\Delta_{\text{w}}}\right)$, where $w_0=W/2$ and $\Delta_{\text{w}}$ are the position and width of the domain wall, respectively, $p_v$ is the circulation number, which takes the value $\pm1$ depending on whether the wall is "down-up" or "up-down", and $p_u$ is the polarity number, which amounts to $\pm1$ depending on whether $\mathbf{M}_{\text{eq}}$ points along $+u$ or $-u$ at the domain wall center. As expected, we find that the spin-wave normal mode of lowest frequency in this configuration is a mode bound to the domain wall [Fig.~\ref{Fig_DW_Profile}(d)], whereas the next one, lying at much higher frequency ($\omega/2\pi\!>\!20$~GHz, see Fig.~\ref{Fig_DW_Profile}(c)), is a bulk-like mode with maximum amplitude near the center of the magnetic domains and zero amplitude at the domain wall location [Fig.~\ref{Fig_DW_Profile}(e)].

\begin{figure}[t]
\includegraphics[width=8.6cm,trim=0 85 0 60,clip]{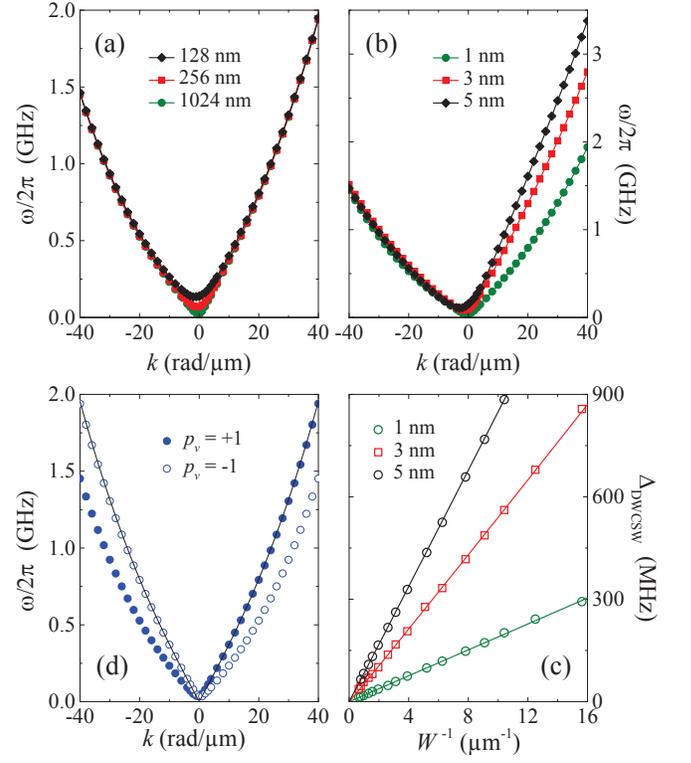}
\caption{(a),(b) Dispersion relations of the DWCSW modes bound to single Bloch walls with $p_u\!=\!p_v\!=\!+1$ in magnetic strips with different width $W$ (a) and thickness $b$ (b). The values of the varying parameter are indicated in the legends. In (a), the thickness is $b\!=\!1$~nm. In (b), the width is $W\!=\!512$~nm. (c) Variation of the frequency gap $\Delta_{\text{DWCSW}}$ with the width of the strip $W$ for different thickness $b$. The lines are linear fits. (d) Dispersion relations of the DWCSW modes bound to single Bloch walls with the same polarity but opposite circulations, in a 1~nm thick 512~nm wide strip: solid (resp. open) symbols correspond to $p_v\!=\!+1$ (resp. $p_v\!=\!-1$). The line is the prediction of Garcia-Sanchez \textit{et al}. [Ref.~\onlinecite{GBSA15}], see text for details.}
\label{Fig_DW_Dispersion}
\end{figure}

In an infinite defect free magnetic medium, the dispersion relation of the DWCSW mode bound to a unique domain wall is gapless since the energy cost of moving the wall as a block, which is what the DWCSW mode with $k\!=\!0$ is all about, is zero\,\cite{LKT09}. Here, a gap is observed whose size, $\Delta_{\text{DWCSW}}$, increases with decreasing strip width $W$ [Fig.~\ref{Fig_DW_Dispersion}(a)] and increasing strip thickness $b$ [Fig.~\ref{Fig_DW_Dispersion}(b)]. This gap is a measure of the restoring force that brings the domain wall back to its equilibrium position in case it is shifted sidewards, which increases dipolar energy. As demonstrated by the scaling of $\Delta_{\text{DWCSW}}$ with the inverse of $W$ [Fig.~\ref{Fig_DW_Dispersion}(c)], the opening of the gap is a finite size effect, which may be viewed as originating from the interaction between the domain wall and the lateral edges of the magnetic medium. We note in passing that the data in Fig.~\ref{Fig_DW_Dispersion}(c) prove the ability of our numerical method to determine accurately normal mode frequencies as small as 20~MHz.

From Fig.~\ref{Fig_DW_Dispersion}(a,b), it is clear that the dispersion relation of the DWCSW mode bound to a Bloch wall is not symmetric about $k\!=\!0$, even in a magnetic medium of thickness as small as 1~nm. The degree of asymmetry at large $k$ increases with increasing thickness but is independent of the strip width. This suggests that the asymmetry is intrinsic in the sense that its source is localized within the domain wall region. Interestingly, also, the dispersion curve is transformed into its symmetric about the frequency axis when the circulation $p_v$ is changed [Fig.~\ref{Fig_DW_Dispersion}(d)], but it is unaffected when the polarity $p_u$ is reversed. This shows that the asymmetry of the dispersion curve is not directly linked to the domain wall chirality, since the latter obeys the same symmetry rules as $p_u\times p_v$. Finally, we note that, as $b$ approches zero, only one of the two branches of the computed $\omega(k)$ curve, either the positive $k$-branch or the negative $k$-branch depending on $p_v$, follows quite closely the relation $\omega(k)=\sqrt{\omega_k(\omega_k+\omega_{\perp})}$, with $\omega_k\!=\!2|\gamma|Ak^2/M_{\text{S}}$ and $\omega_{\perp}=|\gamma|\mu_0N_wM_{\text{S}}$, derived in Ref.~\onlinecite{GBSA15} by treating the domain wall as a magnetic object with effective demagnetization factor $N_w=b/(b+\pi\Delta_{\text{w}})$ along the perpendicular-to-wall axis $w$. Whether this is a coincidence or whether there are good physical reasons for that is a question left to future investigations.

Here, unlike in other works\,\cite{GBSA15}, no interaction that produces a chiral symmetry breaking, like the Dzyaloshinskii-Moriya interaction, is considered. Therefore, the non-reciprocal character of the spin-wave propagation must originate from dipole-dipole interactions, as in the case of surface waves [Sec.~\ref{Sec_Films}]. As a matter of fact, there exists a rather strong similitude between a perpendicularly magnetized strip (with or without a Bloch wall) and a transversally magnetized film. In both cases, indeed, the medium is magnetized in such a way that $\mathbf{M}_{\text{eq}}$ has a (large) component in the plane perpendicular to the direction of spin-wave propagation, $u$, and, conversely, $\mathbf{m}$ possesses a non-zero $u$-component. This is the first necessary ingredient for observing dipole-induced non-reciprocity since, as an examination of the mutual demagnetizing tensors [Eqs.~\ref{Eq_Slab_n_Mutual} and \ref{Eq_Rect_n}] reveals, no dipolar coupling depending on the sign of $k$ can ever exist if $m_u\!=\!0$. For frequency non-reciprocity to occur, a second ingredient is necessary: the magnetic system must not be mirror symmetric about its midplane normal to $\mathbf{e}_u\!\times\!\mathbf{e}_z$ [Ref.~\onlinecite{C87}]. If it is mirror symmetric, non-reciprocal dipolar couplings may play a significant role (they produce asymmetric modal profile in the SW configuration) but they cannot yield any difference in the frequency of counter-propagating spin-waves as their average effect is quantitatively the same for both positive and negative $k$. Here, it is the very presence of the Bloch wall in the strip which breaks the left/right symmetry about the midplane normal to $\mathbf{e}_u\times\mathbf{e}_z\!=\!\mathbf{e}_w$. With the wall sitting at the centre of the strip, there exist no symmetry operation which changes $k$ into $-k$ while leaving the equilibrium magnetic configuration unchanged. To some extent, the presence of the wall is equivalent to having $M_{\text{S}}\!>\!0$ in one half of the strip and $M_{\text{S}}\!<\!0$ in the other.

\begin{figure}[t]
\includegraphics[width=8.5cm,trim=23 75 20 80,clip]{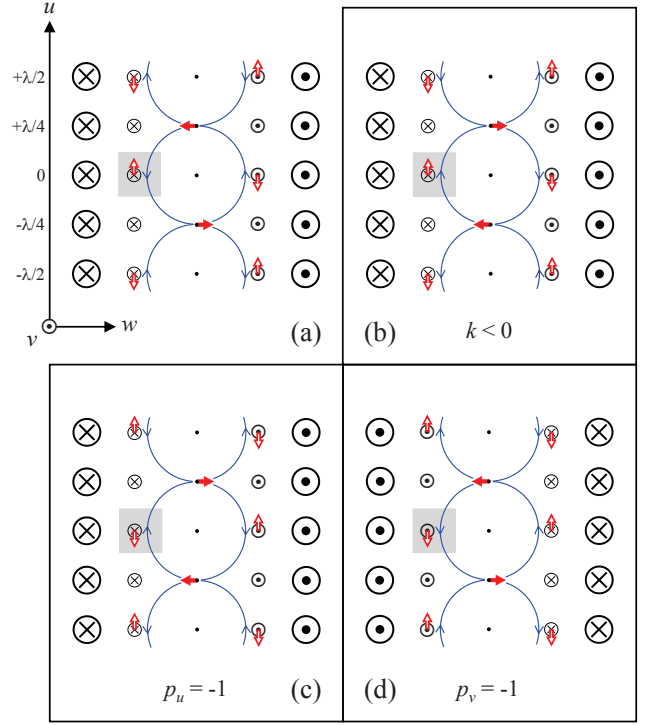}
\caption{Schematic representation of the DWCSW mode hosted by a Bloch wall along a full spin-wave wavelength. The black arrows pointing in and out of the figure represent the out-of-plane component of the equilibrium magnetization $\mathbf{M}_{\text{eq}}$. The open and solid red arrows show the $u$ and $w$-components of the dynamic magnetization, $\mathbf{m}_u$ and $\mathbf{m}_w$, respectively, as deduced from the numerically determined normal modes. The solid blue arrows represent the dipolar field created by $\mathbf{m}_w$ and acting on $\mathbf{m}_u$. (a) Reference case with $p_u\!=\!p_v\!=\!+1$ and $k\!>\!0$. (b) Reversed direction of propagation ($k\!<\!0$). (c) Reversed polarity ($p_u\!=\!-1$). (d) Reversed circulation ($p_v\!=\!-1$). In all four cases, time $t$ is such that $\mathbf{m}_v$ (not shown) points out of the figure at the center of the wall (marked with a black dot), in $u\!=\!0$.}
\label{Fig_DW_Non_Reciprocity}
\end{figure}

Let us examine in detail how non-reciprocal dipolar couplings are affected when either the circulation or the polarity of the wall is changed. For this, we refer to Fig.~\ref{Fig_DW_Non_Reciprocity} where the essential features of the DWCSW mode [Fig.~\ref{Fig_DW_Profile}(d)], as deduced from numerical simulations, are sketched : $\mathbf{m}_u$ and $\mathbf{m}_w$ oscillate in quadrature; $m_{0w}$ is maximum at the center of the wall whereas $m_{0u}$ shows two maxima of opposite signs located symmetrically on either side of the wall center; $\mathbf{m}_v$ plays no decisive role. In this figure, one sees that the dipolar field $\mathbf{h}_{\text{d}u}^{\;\;w}$ (blue arrows) created by $\mathbf{m}_w$ (solid red arrows) and acting on $\mathbf{m}_u$ (open red arrows) reverses when the direction of propagation is reversed [Fig.~\ref{Fig_DW_Non_Reciprocity}(b)]. This is the essence of the non-reciprocity phenomenon, which is reflected in the change of sign of $n_{uw}$ on reversing $\mathbf{k}$ [Eq.~\ref{Eq_Rect_nwu2}]. One also sees that, as far as dynamic dipolar interactions are concerned, changing the polarity of the wall [Fig.~\ref{Fig_DW_Non_Reciprocity}(c)] has no effect since the relative orientation of $\mathbf{m}_u$ and $\mathbf{h}_{\text{d}u}^{\;\;w}$ remains the same, whereas changing the circulation [Fig.~\ref{Fig_DW_Non_Reciprocity}(d)] is equivalent to reversing the direction of propagation (see grey boxes). This explains why the dispersion curves for $p_u\!=\!+1$ and $p_u\!=\!-1$ are identical and those for $p_v\!=\!+1$ and $p_v\!=\!-1$ are symmetric to each other [Fig.~\ref{Fig_DW_Dispersion}(d)].

\section{Possible extensions of the model\label{Sec_Extensions}}
A first possible extension of the model described in this paper would consist in implementing more general and accurate boundary conditions\,\cite{MD07}. Here, for the sake of simplicity, we have assumed so called free boundary conditions, which arise from the sole symmetry breaking of the exchange interactions at the surfaces of the magnetic medium. Moreover, as written in Eqs.~\ref{Eq_Hex2} and \ref{Eq_hex1}, these conditions ($\partial\mathbf{M}_{\text{eq}}/ \partial\mathbf{n}=0$ and $\partial\mathbf{m}/\partial\mathbf{n}=0$, where $\mathbf{n}$ is the normal to the surface) are implemented in the crudest possible way: instead of using accuracy preserving expansions of the spatial derivatives for magnetic cells sitting at or close to the surfaces\,\cite{MD07}, we simply forget altogether, in the expressions of the static and dynamic exchange fields based on second-order Taylor expansions, those pair-interaction-like terms of the form $\frac{\Lambda^2}{\xi^2} (\mathbf{M}_{\text{eq}}^{(\alpha\!-\!1)}\!- \!\mathbf{M}_{\text{eq}}^{(\alpha)})$ or $\frac{\Lambda^2}{\xi^2} (\mathbf{m}^{(\alpha\!-\!1)}\!- \!\mathbf{m}^{(\alpha)})$ which involve missing magnetic cells ($\alpha\!\leqslant\!1$ or $\alpha\!>\!N$). We wish to emphasize however that, with small enough cells, this crude approximation has very little influence on the computed mode profiles and usually none on the frequencies.

Even with free boundary conditions, a number of surface phenomena not discussed above may be included in the model, especially in the film geometry. Surface anisotropies may be introduced as bulk anisotropies present only in the magnetic cells sitting next to the top and/or bottom surfaces. For small enough cells, this is quite equivalent to introducing them through proper boundary conditions. Similarly, an interfacial Dzyaloshinskii-Moriya (DM) interaction, as resulting from a perpendicular-to-plane symmetry breaking\,\cite{BR01}, may also be included. To do so only requires to introduce a new contribution to the dynamic magnetic field experienced by the cell(s) sitting next to the surface(s) since the DM interaction does not contribute to the static effective field $\mathbf{H}_{\text{eq}}$ under the assumption that the orientation of the equilibrium magnetization depends only on the $v$-coordinate. Starting from the expression of the DM energy density given in Eq.~2 of Ref.~\onlinecite{MSLK13} and taking into account the plane wave nature of the spin-waves [Eq.~\ref{Eq_PWSW}], one easily shows that this new contribution has the form $\mathbf{h}_{\text{DM}}^{(\alpha)} = -i\,\frac{2D^{(\alpha)}}{\mu_0M_{\text{S}}^2}\,k \left(\mathbf{e}_w\times\mathbf{m}^{(\alpha)}\right)$, where $D^{(\alpha)}$ is the continuous effective DM constant, in J/m$^2$, possibly different at the top ($\alpha\!=\!N$) and bottom ($\alpha\!=\!1$) surfaces. In the $xyz$ coordinate system, this yields
\begin{subequations}\label{Eq_hDM2}
\begin{align}
(\alpha = 1,N)
\quad
\mathbf{e}_x\cdot\mathbf{h}_{\text{DM}}^{(\alpha)} &= i\,\frac{2D^{(\alpha)}}{\mu_0M_{\text{S}}^2}\,k\,w_z^{(\alpha)}\,m_y^{(\alpha)}
\\
\mathbf{e}_y\cdot\mathbf{h}_{\text{DM}}^{(\alpha)} &= -i\,\frac{2D^{(\alpha)}}{\mu_0M_{\text{S}}^2}\,k\,w_z^{(\alpha)}\,m_x^{(\alpha)},
\end{align}
\end{subequations}
with $w_z^{(\alpha)}\!=\!T_{33}^{(\alpha)}$ the $z$-coordinate of unit vector $\mathbf{e}_w$. We note that introducing the same DM interaction in the strip geometry can only be achieved by simultaneously adding a new contribution to $\mathbf{H}_{\text{eq}}$ and implementing specific exchange-DM boundary conditions at the strip edges\,\cite{RT13,VLDG14,K14}.

Taking into account magnetic damping is another possible extension of the method. With a damping torque of the form proposed by Gilbert, i.e., $\frac{\alpha_\ast}{M_{\text{S}}}(\mathbf{M}\times\dot{\mathbf{M}})$ (damping constant $\alpha_\ast$), equation~\ref{Eq_LLxyz} becomes
\begin{widetext}
\begin{equation}
\omega \left(\begin{matrix} m_x^{(\alpha)}\\ m_y^{(\alpha)} \end{matrix}\right) = \frac{-i|\gamma|\mu_0}{1+\alpha_\ast^2} \left(\begin{matrix} \;\;\;M_{\text{S}}(h_y^{(\alpha)}\!+\!\alpha_\ast h_x^{(\alpha)}) &-& H_{\text{eq}}^{(\alpha)}\,(m_y^{(\alpha)}\!+\!\alpha_\ast m_y^{(\alpha)}) \\ -M_{\text{S}}(h_x^{(\alpha)}\!-\!\alpha_\ast h_y^{(\alpha)}) &+& H_{\text{eq}}^{(\alpha)}\,(m_x^{(\alpha)}\!-\!\alpha_\ast m_y^{(\alpha)}) \\ \end{matrix}\right).
\label{Eq_LLGxyz}
\end{equation}
\end{widetext}
This shows that the construction of the dynamic matrix does not require to evaluate new quantities, just to arrange those considered in the present work in a slightly different manner. With damping included, the eigenfrequencies become complex numbers and their imaginary parts are the inverses of the relaxation times ($T_2$) of the normal modes. Together with the group velocity $v_{\text{g}}$ derived from the dispersion relation, $T_2$ yields the attenuation length $L_{\text{att}}\!=\! v_{\text{g}}T_2$ of a spin-wave mode, which is a parameter of great interest in magnonics.

Finally, moving from a one-dimensional to a two-dimensional array of parallelepipedic cells would be the ultimate extension. It would allow one to describe more accurately what happens in thick strips where the magnetic configuration and/or properties are also inhomogeneous through the thickness of the medium, not just across its width. In practice, this would essentially require to take into account not just two but four nearest neighboring cells in the expressions of the static and dynamic exchange fields.

\section{Conclusion\label{Sec_Conclusion}}
The full recipe has been given for a finite-difference numerical scheme dedicated to the determination of the normal modes of spin-waves propagating as plane-waves in extended magnetic films and strips, in the linear regime. The approach, based on the dynamic matrix method, heavily relies on the use of plane-wave (dynamic) demagnetization factors, for which mathematical expressions have been derived. As illustrated through two examples in the paper, it is well suited to study magnetic media whose material parameters vary in space, like multilayered films, or contain non-collinear micromagnetic textures such as magnetic domain walls. It would allow one exploring spin-wave physics in very complex systems, which are doubly inhomogeneous (both in their magnetic parameters and in their equilibrium magnetic configuration) like, for instance, thin-film hard-soft exchange-spring magnets where planar domain walls can be formed\,\cite{MHHM08}.

The main limitation of the presented micromagnetic model resides in the assumption that the equilibrium magnetic configuration is invariant along the direction of spin-wave propagation. This makes the model unsuitable for studying how spin-waves propagate in the presence of complex magnetic microstructures which never fulfil this condition, like crossties, vortices, or skyrmions. In such situations, one would have to resort to usual time-domain micromagnetic simulations or to other recently developed specific methods\,\cite{BFK14}. We believe that this limitation is amply counterbalanced by the wealth of accurate information that can easily be obtained in situations where the model is applicable, which includes the spatial profiles, frequencies, and dispersion relations of virtually all the propagating spin-wave modes. Besides, a way has been outlined to obtain yet even more micromagnetic information about these modes, by accounting for the effect of magnetic damping and thereby getting access to their relaxation time and attenuation length.

\appendix

\section{Parameters describing the precessional motion of magnetization}\label{App_Precessional_Motion}
\begin{figure}[t]
\includegraphics[width=8.0cm,trim=0 25 0 40,clip]{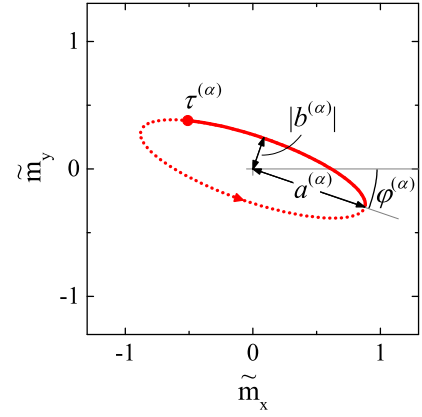}
\caption{Time trajectory of the magnetization for $m_{0x}^{(\alpha)}= -0.51\!+\!0.72\,i$, $m_{0y}^{(\alpha)}=0.38\!-\!0.07\,i$ ($u\!=\!0$, $\omega\!>\!0$), which correspond to $a^{(\alpha)}\!=\!0.93$, $b^{(\alpha)}\!=\!0.26$, $\varphi^{(\alpha)}\!=\!-0.33$~rad, $\tau^{(\alpha)}\!=\!2.28$~rad. The red solid circle marks the position of the magnetization vector at $t\!=\!0$ and the arrow indicates the direction of precession, which is determined by the sign of $\omega a^{(\alpha)}/b^{(\alpha)}$.}
\label{Fig_Trajectory}
\end{figure}

Hereafter, we give mathematical expressions for the four practical parameters that best describe the precessional motion of the magnetization in a given magnetic cell $\alpha$ as a function of the not-so-convenient complex amplitudes $m_{0x}^{(\alpha)}$ and $m_{0y}^{(\alpha)}$. Assuming $u\!=\!0$, the time trajectory of the true variable magnetization $\tilde{\mathbf{m}}^{(\alpha)}= \text{Re}(\mathbf{m}^{(\alpha)})$ of cell $\alpha$, in the ($x$,$y$) plane, is an ellipse [Fig.~\ref{Fig_Trajectory}], whose parametric equations are
\begin{equation}
\left\lbrace
\begin{array}{l}
\tilde{m}_x^{(\alpha)}(t) = \text{Re}(m_{0x}^{(\alpha)})\cos\omega t - \text{Im}(m_{0x}^{(\alpha)})\sin\omega t\\
\tilde{m}_y^{(\alpha)}(t) = \text{Re}(m_{0y}^{(\alpha)})\cos\omega t - \text{Im}(m_{0y}^{(\alpha)})\sin\omega t
\end{array}\right..
\end{equation}
Comparing them to the general form for an ellipse centered at the origin
\begin{equation}
\left\lbrace
\begin{array}{ll}
\tilde{m}_x^{(\alpha)}(t) =& a^{(\alpha)}\cos\!\varphi^{(\alpha)}\cos(\omega t\!+\!\tau^{(\alpha)}) \\ &-b^{(\alpha)}\sin\!\varphi^{(\alpha)}\sin(\omega t\!+\!\tau^{(\alpha)})\\
\tilde{m}_y^{(\alpha)}(t) =& a^{(\alpha)}\sin\!\varphi^{(\alpha)}\cos(\omega t\!+\!\tau^{(\alpha)}) \\ &+b^{(\alpha)}\cos\!\varphi^{(\alpha)}\sin(\omega t\!+\!\tau^{(\alpha)})
\end{array}
\right.
\end{equation}
and introducing the intermediate variables
\begin{subequations}
\begin{align}
\eta_{\pm}^{(\alpha)} &= \text{Re}\!\left(m_{0x}^{(\alpha)}\right) \pm \text{Im}\!\left(m_{0y}^{(\alpha)}\right)
\\
\zeta_{\pm}^{(\alpha)} &= \text{Re}\!\left(m_{0y}^{(\alpha)}\right) \pm \text{Im}\!\left(m_{0x}^{(\alpha)}\right),
\end{align}
\end{subequations}
we find
\begin{subequations}\label{Eq_Ellipse_Semi_Axes}
\begin{align}
a^{(\alpha)} &= \frac{\left(\left|\mathbf{m}_0^{(\alpha)}\right|^2 +\left({\eta_{+}^{(\alpha)}}^2\!+{\zeta_{-}^{(\alpha)}}^2\right)^{\frac{1}{2}} \left({\eta_{-}^{(\alpha)}}^2\!+{\zeta_{+}^{(\alpha)}}^2\right)^{\frac{1}{2}} \right)^{\frac{1}{2}}}{\sqrt{2}}
\nonumber\\ \text{and}
\\ b^{(\alpha)} &= \frac{\left|\mathbf{m}_0^{(\alpha)}\right|^2 - {\eta_{+}^{(\alpha)}}^2 - {\zeta_{-}^{(\alpha)}}^2} {2a^{(\alpha)}},
\end{align}
\end{subequations}
for the ellipse semi-axes $a^{(\alpha)}\!>\!0$ and $b^{(\alpha)}$ ($|b^{(\alpha)}|\!\leqslant\!a^{(\alpha)}$),
\begin{equation}
\varphi^{(\alpha)} =
\frac{1}{2}\left[
\text{Arg}\!\left(\eta_{-}^{(\alpha)}+i\zeta_{+}^{(\alpha)}\right) + \text{Arg}\!\left(\eta_{+}^{(\alpha)}+i\zeta_{-}^{(\alpha)}\right)
\right],
\label{Eq_Ellipse_Tilt_Angle}
\end{equation}
for the tilt angle of the ellipse major axis with respect to the $x$-axis, and 
\begin{align}
\tau^{(\alpha)} =& \left\{\frac{1}{2}\left[ \text{Arg}\!\left(\eta_{-}^{(\alpha)}+i\zeta_{+}^{(\alpha)}\right) - \text{Arg}\!\left(\eta_{+}^{(\alpha)}+i\zeta_{-}^{(\alpha)}\right)\right] \right.\nonumber\\
&\left. +\,2\pi n \quad|\quad n \in \mathbb{Z} \vphantom{\left\{ \text{Arg}\!\left(\eta_{-}^{(\alpha)}+i\zeta_{+}^{(\alpha)}\right) - \text{Arg}\!\left(\eta_{+}^{(\alpha)}+i\zeta_{-}^{(\alpha)}\right)\right.} \right\}.
\label{Eq_Ellipse_Phase}
\end{align}
for the phase of the precessional motion. We note that this set of equations [Eqs.~\ref{Eq_Ellipse_Semi_Axes}-\ref{Eq_Ellipse_Phase}] is not unique and that $\text{Re}(m_{0y}^{(\alpha)})\! = \!\text{Im}(m_{0x}^{(\alpha)})\!=\!0$ implies $\varphi^{(\alpha)}\!=\!\tau^{(\alpha)}\!=\!0$ and vice versa. Also, while a tilt angle outside the range $(-\pi,\pi]$ would bear no physical meaning, $\tau^{(\alpha)}$ can take on values outside this range in order to account for relative changes of phase exceeding $2\pi$ across a mode profile. Such a situation may indeed arise in specific circumstances, for instance, in the case of the DWCSW mode associated with a N\'eel wall in an in-plane magnetized strip.

\section{Static mutual demagnetization factors of rectangular parallelepipeds with infinite length}\label{App_Static_Demag_Tensor}
In this appendix, we derive analytical expressions for the static mutual demagnetization factors between parallelepipedic magnetic cells with infinite length in the $u$-direction and rectangular $(b\!\times\!c)$ cross section in the $(v,w)$ plane. It is assumed that the source cell, which creates the dipolar field, is centered in $(v,w)\!=\!(0,0)$, whereas the target cell, which experiences it, is centered at the relative coordinates $(\delta v,\delta w)$. The starting point of the calculation is the well-known expression of the magnetic field created by a one-dimensional distribution of magnetic charges $\sigma_0\,\delta(v-v_0)\delta(w-w_0)$ with linear density $\sigma_0$, parallel to axis $u$, that is,
\begin{equation}
\mathbf{H}_{\text{1D}}(\sigma_0,\rho) = \frac{\sigma_0}{2\pi\rho}\, \mathbf{e}_{\rho},
\label{Eq_Rect_Magn_Field}
\end{equation}
where $\rho\!=\!\sqrt{(v\!-\!v_0)^2\!+\!(w\!-\!w_0)^2}\,$ is the radial distance to the line of charges and $\mathbf{e}_{\rho}\!= \!\frac{v\!-\!v_0}{\rho}\,\mathbf{e}_v+\frac{w\!-\!w_0}{\rho}\,\mathbf{e}_w$ is a unit vector in the radial direction.

If the source cell is saturated along $w$, surface charges $\pm M_{\text{S}}$ are created on its vertical faces. The stray field produced is obtained by integrating $\mathbf{H}_{\text{1D}}$ over $v_0\!\in\![-\frac{b}{2},+\frac{b}{2}]$, with $\sigma_0\!=\!\pm M_{\text{S}}\,dv_0$ in $w_0\!=\!\pm\frac{c}{2}$. Then two more integrations over $v\!\in\![\delta v-\frac{b}{2},\delta v+\frac{b}{2}]$ and  $w\in[\delta w-\frac{c}{2},\delta w+\frac{c}{2}]$ are necessary to calculate its average value over the volume of the target cell, $\mathbf{H}_{\text{d}}$. Finally, the demagnetization factor $N_{iw}$ (with $i=u,v,w$) may be identified as the factor that makes the $i$-component of $\mathbf{H}_{\text{d}}$ equal to $-N_{iw}M_{\text{S}}$. The full calculation is long but straightforward. It yields
\begin{widetext}
\begin{align}
N_{ww}(\delta v,\delta w) = \frac{1}{2\pi bc}\sum_{n=-1}^{1}\sum_{m=-1}^{1} (2\!-\!3|n|)(2\!-\!3|m|)&\left\{(\delta v+nb)(\delta w+mc)
\,\arctan\!\left(\frac{\delta v+nb}{\delta w+mc}\right)\right. \nonumber\\
&+\left.\frac{(\delta v+nb)^2-(\delta w+mc)^2}{4}
\,\ln\!\left[(\delta v+nb)^2+(\delta w+mc)^2\right]\right\}
\label{Eq_Static_Nww}
\end{align}
and
\begin{align}
N_{vw}(\delta v,\delta w) = \frac{1}{4\pi bc}\sum_{n=-1}^{1}\sum_{m=-1}^{1}
(2\!-\!3|n|)(2\!-\!3|m|)&\left\{(\delta v+nb)^2
\,\arctan\!\left(\frac{\delta w+mc}{\delta v+nb}\right)\right. \nonumber\\
&+(\delta v+nb)(\delta w+mc) \,\ln\!\left[(\delta v+nb)^2+(\delta w+mc)^2\right]\frac{}{}
\nonumber\\
&+\left.(\delta w+mc)^2\,\arctan\!\left(\frac{\delta v+nb}{\delta w+mc}\right)\right\}.
\label{Eq_Static_Nvw}
\end{align}
\end{widetext}
$N_{uw}$ and, more generally, all $N_{ui}$ elements ($i=u,v,w$) are nil since $\mathbf{H}_{\text{1D}}$ has no component along $u$. From Eq.~\ref{Eq_Static_Nvw}, one may see that $N_{vw}$ is also nil as soon as either $\delta v$ or $\delta w$ is zero.

The $N_{iv}$ elements can be calculated in a similar manner, by assuming that the source cell is saturated along $v$ and that surface charges $\pm M_{\text{S}}$ are therefore created on its horizontal faces $v_0\!=\!\pm\frac{b}{2}$. Alternatively, they can also be deduced by using the intrinsic properties of the demagnetizing tensor, namely the fact that it is symmetric and that its trace equals the fraction of the volume of the source cell which overlaps that of the target cell\,\cite{NWD93}. For totally disjoint rectangular parallepipeds with infinite length, this means $N_{iu}=N_{ui}=0$ ($i=u,v,w$), $N_{wv}=N_{vw}$, and $N_{vv}=-N_{ww}$.

\section{Magnetic field from a one-dimensional harmonic distribution of magnetic charges.}\label{App_Magn_Field_From Harmonic_Dist}
Our goal here is to derive an analytical expression for the magnetic field $\mathbf{h}_{\text{1D}}$ created by a one-dimensional harmonic distribution of magnetic charges parallel to axis $u$ and located at the transverse position $(v_0,w_0)$, as defined by Eq.~\ref{Eq_1D_Charge_Dist}. We start by looking for the corresponding magnetostatic potential $\phi_{\text{1D}}$, which obeys Laplace's equation $\Delta\phi_{\text{1D}}=0\;$ everywhere in space but at the position of the line of charges. To solve this problem, cylindrical coordinates $(\rho,\theta,u)$ are more appropriate than the cartesian coordinates $(u,v,w)$. Moreover, the charge distribution is such that the solution is expected to be of the form
\begin{equation}
\phi_{\text{1D}}(\rho,u,t)=\tilde{\phi}_{\text{1D}}(\rho)\,e^{i(\omega t-ku)},
\label{Eq_Rect_Magn_Potential_1}
\end{equation}
where $\rho\!=\!\sqrt{(v\!-\!v_0)^2\!+\!(w\!-\!w_0)^2}\,$ is once again the radial distance to the line of charges. Introducing this trial solution into Laplace's equation and performing the change of variable $\epsilon=k\rho$, we find that $\tilde{\phi}_{\text{1D}}$ must obey
\begin{equation}
\epsilon^2\frac{\partial^2\tilde{\phi}_{\text{1D}}}{\partial\epsilon^2}+\epsilon \frac{\partial\tilde{\phi}_{\text{1D}}}{\partial\epsilon} - \epsilon^2\tilde{\phi}_{\text{1D}} = 0.
\label{Eq_Rect_Magn_Potential_2}
\end{equation}
General solutions to Eq.~\ref{Eq_Rect_Magn_Potential_2} are linear combinations of the zero-th order modified Bessel functions of the first $(I_0)$ and second $(K_0)$ kinds. However, $I_0$ cannot be part of a physical solution since it diverges when its argument goes to both positive infinity and negative infinity. As for $K_0$, it takes on complex values for negative real arguments and diverges at negative infinity. Then the magnetostatic potential $\phi_{\text{1D}}$ must be of the form
\begin{equation}
\phi_{\text{1D}}(\rho,u,t)=A\,K_0(|k|\rho)\,e^{i(\omega t-ku)}
\label{Eq_Rect_Magn_Potential_3}
\end{equation}
and the magnetic field deriving from it must write
\begin{align}
\mathbf{h}_{\text{1D}}(\rho,u,t) =& -\!\boldsymbol\nabla\phi_{\text{1D}}(\rho,u,t)
\nonumber\\
=&\,A\,k\,e^{i(\omega t-ku)}
\nonumber\\
&\times \left[\,i\,K_0(\!|k|\rho)\,\mathbf{e}_u + \text{sgn}(k)\,K_1(|k|\rho)\,\mathbf{e}_{\rho}\,\right],
\label{Eq_Rect_Magn_Field}
\end{align}
where $K_1(\epsilon)=-\frac{\partial K_0(\epsilon)}{\partial\epsilon}$ is the first-order modified Bessel function of the second kind and $\mathbf{e}_{\rho}\!=\!\frac{v\!-\!v_0}{\rho}\,\mathbf{e}_v+ \frac{w\!-\!w_0}{\rho}\,\mathbf{e}_w$, as before.

To determine the unknown prefactor $A$, we may use Gauss's theorem. To this end, we construct a Gauss volume consisting of a cylinder of radius $R$, length $L$, and axis merged with the line of magnetic charges. This cylinder is bounded by three surfaces: The two circular end surfaces denoted $\Sigma_1$ and $\Sigma_3$, and the lateral surface called $\Sigma_2$. If we make the cylinder radius tend to zero, the flux of $\mathbf{h}_{\text{1D}}$ through $\Sigma_1$ and $\Sigma_3$ vanishes because $\lim\limits_{R\,\to\,0}\left[|k|R\,K_0(|k|R) \right]=0$, whereas the flux of $\mathbf{h}_{\text{1D}}$ through $\Sigma_2$ is
\begin{align}
\Phi_{\Sigma_2}=&\lim\limits_{R\,\to\,0}\oiint\limits_{\Sigma_2} \left(\mathbf{e}_{\rho}\cdot\mathbf{h}_{\text{1D}}\right)d\Sigma_2
\nonumber\\
=&\lim\limits_{R\,\to\,0}\int_{u}^{u+L}\!\!\int_{0}^{2\pi}\! A|k|R\,K_1\!(|k|R)\,e^{i(\omega t-ku)}\,d\theta\,du
\nonumber\\
=&\;2\pi A \int_{u}^{u+L}\!e^{i(\omega t-ku)}\,du,
\end{align}
using $\lim\limits_{R\,\to\,0}\left[|k|R\,K_1(|k|R)\right]=1$. Equating $\Phi_{\Sigma_2}$ with the total magnetic charge contained in the cylinder
\begin{equation}
Q_\text{M} = \sigma_0\int_{u}^{u+L}\!e^{i(\omega t-ku)}\,du,
\end{equation}
we readily find
\begin{equation}
A = \frac{\sigma_0}{2\pi}.
\label{Eq_Prefactor}
\end{equation}

\section{Test of the plane-wave demagnetizing tensor approach}\label{App_Validation}
In order to demonstrate the correctness of our theoretical results concerning the plane-wave demagnetizing tensor of magnetic cells having the shape of extended slabs [Sec.~\ref{Sec_Slabs}], normal mode profiles computed for homogeneous extended films in the purely magnetostatic (exchange-free) case have been compared to predictions of the exact analytical model developed by Damon and Eshbach\,\cite{DE61,SP09}. For the lowest-order even backward volume wave mode ($\mathbf{e}_z=\!\mathbf{e}_u$), this model adapted to our geometry [Fig.~\ref{Fig_Magnetic_Cells}(a)] and conventions predicts
\begin{subequations}\label{Eq_Magnetostatic_BVW}
\begin{align}
m_{0v} &= -i\,\phi_0\,\nu\,\chi\,k_v\,\cos\!\left(k_v\left(v\!-\!T/2\right)\right) \label{Eq_Magnetostatic_BVW_v}\\
m_{0w} &= -\;\phi_0\,\kappa\,k_v\,\cos\!\left(k_v\left(v\!-\!T/2\right)\right) ,\label{Eq_Magnetostatic_BVW_w}
\end{align}
\end{subequations}
where $\phi_0$ is a constant which depends on the normalization conditions, $\nu\!=\!\text{sgn}(k)$, and
\begin{equation}
\chi\!=\!\frac{\omega_{\text{M}}\omega_{\text{H}}} {\omega_{\text{M}}^2-\omega^2},\quad \kappa\!=\!\frac{\omega_{\text{M}}\omega}{\omega_{\text{M}}^2-\omega^2},\quad k_v\!=\!\frac{-k^2}{1+\chi},
\end{equation}
with $\omega_{\text{M}}\!=\!|\gamma|\mu_0M_{\text{S}}$ and $\omega_{\text{H}}\!= \!|\gamma|\mu_0H_0$. For the surface wave mode ($\mathbf{e}_z\!=\!\mathbf{e}_w$), the model yields
\begin{subequations}\label{Eq_Magnetostatic_SW}
\begin{align}
m_{0u} &= \;\phi_0\,|k|\,\frac{\left(\nu\chi-\kappa\right)e^{|k|v}+
p(\nu)\left(\nu\chi+\kappa\right)e^{-|k|v}}{A(\nu)}
\label{Eq_Magnetostatic_SW_u}\\
m_{0v} &= i\,\phi_0|k|\,\frac{\left(\nu\chi-\kappa\right)e^{|k|v}- p(\nu)\left(\nu\chi+\kappa\right)e^{-|k|v}}{A(\nu)}
\label{Eq_Magnetostatic_SW_v},
\end{align}
\end{subequations}
where
\begin{equation}
p(\nu) = \frac{\chi+2-\nu\kappa}{\chi+\nu\kappa}
\end{equation}
and
\begin{equation}
A(\nu) = \begin{cases}
2\,e^{|k|T/2} &\nu = 1\\
\,e^{-|k|T/2}\left[\left(\chi+2+\kappa\right)e^{2|k|T}\!-\!\left(\chi+\kappa\right)\right] &\nu = -1
\end{cases}.
\end{equation}

\begin{figure}[t]
\includegraphics[width=8.5cm,trim=45 55 50 48,clip]{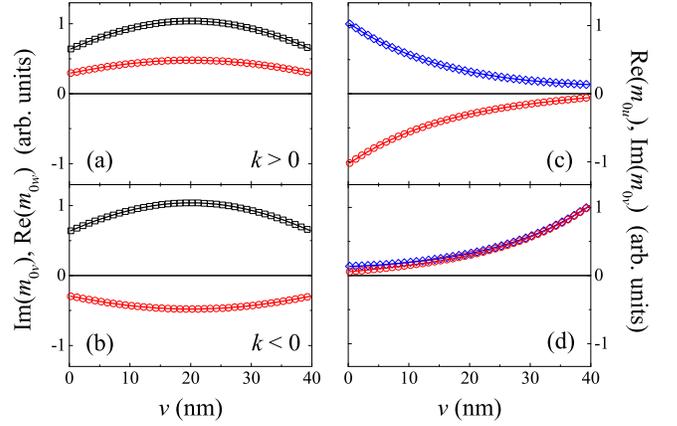}
\caption{Profiles of the lowest-order even backward volume wave mode (a,b) and surface wave mode (c,d) with $|k|\!=\!60$~rad/$\mu$m, in a 40~nm thick film with $A\!=\!0$ and $M_{\text{S}}\!=\!800$~kA/m ($\mu_0H_0\!=\!0.1$~T). Symbols and lines correspond to results of our numerical approach ($b=0.5$~nm) and predictions of the Damon-Eshbach analytical model [Eqs.~\ref{Eq_Magnetostatic_BVW} and \ref{Eq_Magnetostatic_SW}], respectively. The complex amplitudes $m_{0u}$ (blue diamonds), $m_{0v}$ (red circles), and $m_{0w}$ (black squares) are shown for both $k>0$ (a,c) and $k<0$ (b,d).} \label{Fig_DE}
\end{figure}

Figure~\ref{Fig_DE} shows the mode profiles calculated using the above two sets of analytical expressions (lines), Eqs.~\ref{Eq_Magnetostatic_BVW} and \ref{Eq_Magnetostatic_SW}, together with results of our numerical approach (symbols), for a particular wave vector value $|k|\!=\!60$~rad/$\mu$m. The two types of data match each other perfectly, for both surface and volume waves. Since such a test is rather demanding, we may conclude that dipole-dipole interactions are correctly accounted for by using the dynamic demagnetizing tensors derived in Sec.~\ref{Sec_Slabs}.

\begin{figure}[t]
\includegraphics[width=8.7cm,trim=30 48 30 25,clip]{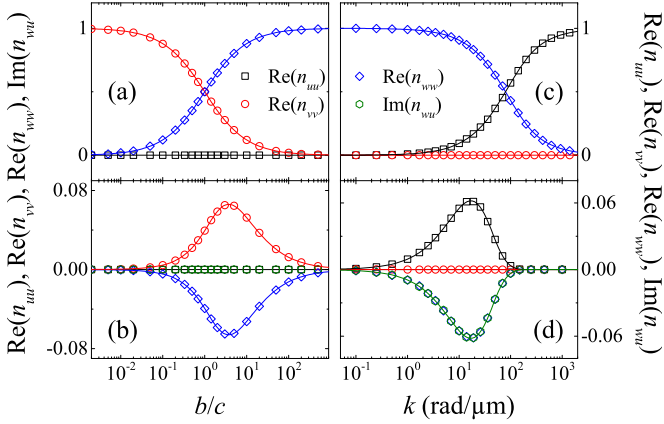}
\caption{(a) Self and (b) mutual ($w_{\alpha}-w_{\beta}=+2c$) dynamic demagnetization factors of parallelepipedic magnetic cells versus cell aspect ratio ($b=10$~nm), as calculated numerically for $k=10^{-12}$~rad/$\mu$m. (c) Self and (d) mutual ($w_{\alpha}-w_{\beta}=-3c$) dynamic demagnetization factors of parallelepipedic cells versus wave vector, as calculated numerically for $b\!=\!200~\mu$m and $c\!=\!20$~nm. Data computed using the integral expressions derived in Sec.~\ref{Sec_Parallelepipeds} (symbols) are compared to analytical results (lines) for (a,b) $k=0$ [Eqs.~\ref{Eq_Trace_N}-\ref{Eq_Rect_Nww_Mut}] and (c,d) $b\rightarrow+\infty$ [Eqs.~\ref{Eq_Slab_nuu_Self}, \ref{Eq_Slab_nuu_Mutual}, and \ref{Eq_Slab_nuv_Mutual}, with $b$ and $v$ replaced with $c$ and $w$].} \label{Fig_Test_Rect}
\end{figure}

In the case of parallelepipedic magnetic cells [Sec.~\ref{Sec_Strips}] and of spin-wave medium having the shape of a strip, such demanding tests as reported above for films could not be performed since fully analytical theories are not available, which could be used for comparison. The only tests we could devise consist in examining limiting cases. A first natural test is to check that the dynamic demagnetization factors defined by the integral expressions Eqs.~\ref{Eq_Rect_nuu2}, \ref{Eq_Rect_nww2}, and \ref{Eq_Rect_nwu2} behave properly when the wave vector $k$ tends to zero. Figure \ref{Fig_Test_Rect}(a,b) shows that this is indeed the case: all factors become equal to their static counterparts given by  Eqs.~\ref{Eq_Trace_N}\,-\ref{Eq_Rect_Nww_Mut}. Another possibility is to investigate what happens when the height $b$ of the cells becomes much larger than both the cell width $c$ and the spin-wave wavelength $\lambda\!=\!2\pi/|k|$. Figure~\ref{Fig_Test_Rect}(c,d) shows that, as expected, the dynamic demagnetization factors are then very close to those of extended slabs and obey the analytical expressions derived in Sec.~\ref{Sec_Slabs} (Eqs.~\ref{Eq_Slab_nuu_Self}, \ref{Eq_Slab_nuu_Mutual}, and \ref{Eq_Slab_nuv_Mutual} with $b$ and $v$ replaced with $c$ and $w$, respectively). Although these two tests are not as stringent as those performed for the film geometry, they support our claim that dipolar interactions can also be well described by dynamic demagnetization factors in the strip geometry.
\begin{equation}\nonumber\end{equation}

\begin{acknowledgments}
The authors thank Joo-Von Kim, Felipe Garcia-Sanchez, Riccardo Hertel, and Andr\'e Thiaville for fruitful discussions, and acknowledge financial support from the French National Research Agency (ANR) under Contract No.~ANR-11-BS10-0003 (NanoSWITI). O.~G. thanks IdeX Unistra for doctoral funding.
\end{acknowledgments}


\begin{thebibliography}{0}
\bibitem{Rem1}
Eigenmodes in a pair have eigenfrequencies with different abolute values only in specific circumstances where so-called frequency non-reciprocity effects take place. See Sec.~\ref{Sec_Applications}.
\bibitem{Rem2}
When the magnetization precesses in phase all along a mode profile, it is always possible to make the condition $\tau^{(\alpha)}\!=\!0$ true for all $\alpha$. Should it not be automatically fulfilled by the numerical routine used for diagonalizing the dynamic matrix, it can always be realized afterwards, by multiplying all complex amplitudes $m_{0x}^{(\alpha)}$ and $m_{0y}^{(\alpha)}$ by an appropriate $e^{i\theta}$ factor. This does not change the physical nature of the eigenmodes since these are defined up to a constant scalar multiplier. Making $\varphi^{(\alpha)}\!=\!0$, on the other hand, is a matter of choosing adequate local reference frames (which are free to rotate about axis $z$) such that axis $x$ always corresponds to the major axis of the elliptical trajectory followed by $\mathbf{M}^{(\alpha)}$. For that, a good choice is usually to have axis $x$ lying in the plane of the magnetic medium.
\bibitem{Rem3}
Equation~\ref{Eq_ha1} can be readily obtained by replacing $\mathbf{M}_{\text{eq}}$ with $(\mathbf{M}_{\text{eq}}+\mathbf{m})$ in Eq.~\ref{Eq_Ha} and retaining all terms linear in $\mathbf{m}$.
\bibitem{Rem5}
In this paper, we abusively call surface waves all spin-waves propagating at right angle to the equilibrium magnetization in an in-plane magnetized medium. However, only the mode with the lowest quantization number along the film   thickness ($n\!=\!0$) has a strong surface character. Modes with $n\!>\!0$ are more bulk-like perpendicular standing spin wave modes.
\end{thebibliography}
\end{document}